# Structural pathway for nucleation and growth of topologically close-packed phase from parent hexagonal crystal

Junyuan Bai[1], Hongbo Xie[1,2], Xueyong Pang[1*], Min Jiang[1,3], Gaowu Qin[1,3*]

[1]*Key Laboratory for Anisotropy and Texture of Materials (Ministry of Education), School of Materials Science and Engineering, Northeastern University, Shenyang 110819, China*

[2]*State Key Laboratory of Rolling and Automation, Northeastern University, Shenyang 110819, China*

[3]*Research Center for Metal Wires, Northeastern University, Shenyang 110819 China*

**Abstract:**

The solid diffusive phase transformation involving the nucleation and growth of one nucleus is universal and frequently employed but has not yet been fully understood at the atomic level. Here, our first-principles calculations reveal a structural formation pathway of a series of topologically close-packed (TCP) phases within the hexagonally close-packed (hcp) matrix. The results show that the nucleation follows a nonclassical nucleation process, and the entire hcp→TCP structural transformation occurs through shuffle-based displacements, with a specific 3-layer hcp-ordering serving as the basic structural transformation unit (BSTU). The thickening of plate-like TCP phases relies on the formation of these hcp-orderings at their coherent TCP/matrix interface to nucleate ledge, but the ledge lacks the dislocation characteristics commonly considered in the conventional view. Furthermore, the atomic structure of the critical nucleus for the $Mg_2Ca$ and $MgZn_2$ Laves phases was predicted in terms of Classical Nucleation Theory (CNT), and the formation of polytypes and off-stoichiometry in TCP precipitates is found to be related to the nonclassical nucleation behavior. These insights contribute to a deeper understanding of solid diffusive phase transformations at the atomic level and provide a foundation for investigating other technologically significant transformations of this kind.

Corresponding author.
Email: pangxueyong@mail.neu.edu.cn (X.Y. Pang)
qingw@smm.neu.edu.cn (G. W. Qin)

## 1. Introduction

Utilizing solid-state phase transformation is widely recognized as an effective approach for designing and fabricating structural and functional materials with tailored properties. To gain a comprehensive understanding of a phase transformation, it is crucial to determine its formation pathway connecting two dissimilar crystallographic structures. However, due to the transient nature of the initial stages in phase transformations, only a few non-diffusive (displacive) transition pathways between some simple crystalline structures have been well documented, such as the Bain pathway[1] (bcc-to-hcp), Burgers pathway[2] (bcc-to-fcc) and bcc-to-$\omega$ pathway [3]. In contrast, most diffusive transition pathways, particularly for transformations from simple structure to complex structure (i.e., simple-to-complex), are still largely unexplored, owing to the high spatial and temporal resolution requirements in the early stages of nucleation and growth pose significant challenges for both experiments and simulations [4–6] Consequently, understanding the nucleation and growth mechanism has long been a persistent challenge in the field of solid-diffusive transformation.

Nucleation is generally regarded as a rare event *via* a stochastic process [5,7]. For this complex process, the Classical Nucleation Theory (CNT) assumes that the nucleus possesses the same crystal structure and composition as the final equilibrium product phase. However, this assumption fails to explain many metastable intermediates observed in numerous solid (liquid)-to-solid phase transformations [5,6,8]. These intermediates significantly differ from the equilibrium phase in terms of crystal structure and composition. To address this common phenomenon, a nonclassical nucleation mechanism that assumes a nucleus has a non-uniform composition and order parameter profile was proposed [9] and has achieved remarkable success in explaining many liquid-to-solid transitions [5,6], as well as a hcp→bcc structural transformation in the Ti-Mo alloy [10]. In addition, it was generally accepted that the formation of solute clusters contributes to the nucleation process [11], which is well understood for isostructural transformations like fcc→$L1_2$ [12]. However, for the more common secondary phase in alloys that involves simple-to-complex structural transformations, the specific role played by these solute clusters still remains elusive.

As for the ensuing growth stage, a ledge-mediated-growth mechanism [7] for the plate-shaped precipitates (our main research subject) in alloys has been built in the traditional solid-state phase transformation theory. Through nucleating ledges (typically considered to have dislocation $\boldsymbol{b}$ and height $\boldsymbol{h}$ characters [13]) on their flat terraces and migrating laterally towards the matrix, the plate-like precipitates evolve in three-dimensional (3D) space. Nevertheless, our understanding of these thickening and lengthening processes remains at the mesoscopic scale, especially the atomic configurations of ledges, as well as their nucleation and migration mechanisms, are still unclear. After all, these commonly observed precipitate plates in alloys, such as the well-known $\theta'$[14], $T_1$[15], Ω phases[16] in aluminum alloys and Laves phases[11] in magnesium alloys, often keep coherence with the surrounding matrix and perform critical technological functions. This makes it crucial to understand the growth mechanism of plate-like precipitates to further improve their thermal stability.

To elucidate the general characteristics of the nucleation-growth mechanism at the atomic scale for a simple-to-complex diffusive transformation, we systematically investigated the formation pathways of a series of topologically-close-packed phases (TCP, also designated as the Frank-Kasper phase[17,18]) from the hcp-matrix, using chemically accurate first-principles calculations. Wherein Mg alloys were selected as our primary model materials because they contain many experimentally observed TCP phases, which can be used as references for our simulations and conversely, to validate our results. Furthermore, TCP precipitates in Mg alloys [19–23] have been found to considerably enhance creep resistance, in addition to providing strength. The other TCP-architectured compounds also exhibit promising applications in various fields, including magnetic [24], electronic [25], and hydrogen storage [26] applications. Here, we focus on revealing the transition pathway of in situ precipitating TCP phases within the hcp-matrix. In this regard, Natarajan et al. [27] first reported that TCP structures could be directly generated through specific hcp-orderings after a structural relaxation in the first-principles calculations. However, their oversimplified models, which ignored matrix constraints, failed to depict the actual nucleation and growth processes (for details see Supplementary texts-I). As a result, the underlying characteristics of hcp→TCP transformations have not been unraveled yet.

By introducing a new crystallographic definition for TCP structures, our study identified a 3-layer unstable hcp-ordering as the basic structural transformation unit (BSTU) of hcp→TCP transformations, and the entire structural transformations, including the lengthening and thickening of TCP plates within the hcp-matrix, is governed by the formation of this BSTU. The nucleation of TCP phases exhibits nonclassical nucleation behavior. Through exploring the kinetic conditions that initiate structural transformations, a basic structural evolution rule of TCP plates in hcp-matrix has been formulated and the atomic structure of the critical nucleus for two common Laves phases was further predicted. Moreover, these findings align well with experimental observations. The insights gained are expected to improve our understanding of solid-diffusive transformations at the atomic level and help explore new strategies for designing novel materials containing TCP phases.

## 2. Methods

## 2.1. First-Principles Methods

First-principles DFT calculations were performed using the Vienna ab initio simulation package (VASP)[28,29] with Blochl's projector augmented wave (PAW) potential method [30]. The exchange-correlation energy functional was described with the generalized gradient approximation (GGA) as parameterized by Perdew-Burke-Ernzerhof (PBE)[31]. The frozen core pseudopotentials were used for RE elements as these pseudopotentials have been shown to replicate thermodynamics and elastic properties of rare-earth intermetallics[32,33]. A 520 eV plane wave cutoff was adopted with convergence criteria for energy and atomic force were set as $10^{-6}$ eV and $10^{-2}$ eV/Å, respectively. Partial occupancies were determined by using the first order Methfessel-Paxton method with a smearing width of 0.2 eV[34]. Relaxations of atomic coordinates and optimizations of the shape and size of the model were adopted for all calculations. A Γ-centered k-point mesh of 18×18×10 was adopted for the HCP primitive cell and other supercells were scaled appropriately to keep k-point density remain constant. Structures are visualized using the VESTA[35].

## 2.2. Employed Models

In this work, we employed the model incorporating the matrix portion to conduct first-principles calculations. The constrained effect of the matrix on the hcp-to-TCP structural transformations is rigorously examined, which significantly differs from the model that excludes the matrix adopted by Natarajan et al. [27]. In Supplementary texts-I, we emphasize that models excluding the matrix component may produce spurious phenomena and yield misleading results in simulations.

The unit-cell structure of $M_R$-type ordering is constructed based on their *a*-vector, *b*-vector, and *c*-vector parallel to the $[10\bar{1}0]_{hcp}$, $[01\bar{1}0]_{hcp}$ and $[0001]_{hcp}$ directions, respectively. Other various-sized models utilized in this work are essentially constructed using this unit-cell structure, such as the 270-atom (3×3×5) models used in Fig. 3 and 450-atom (5×5×3) models used in Figs. A1, A4, and A5. The effects of super-cell sizes on calculation accuracy are presented in Fig. S3.

## 2.3. Energy landscape

The energy landscapes (Fig. 8) show the variation of ($E(\xi_x, \xi_y)$- $E(1, 1)$) as a function of $\xi_x$ and $\xi_y$, where $E(\xi_x, \xi_y)$ denotes the energy of a structure with $\xi_x$ and $\xi_y$ displacements and $E(1, 1)$ signifies the energy of the final relaxed structure. The $E(\xi_x, \xi_y)$ are obtained by interpolating 14×9 intermediate configurations between initial unrelaxed ($\xi_x$=0, $\xi_y$=0) and final relaxed structures, with their internal TCP atoms fixed and minimizing the energy concerning all degrees of freedom. Because a regular

TCP structure does not form instantly but rather gradually, for simplicity, we herein adopt two-dimensionally infinite models (that is, excluding lateral matrix in the model) to plot energy landscapes, and the four situations shown in Fig. 8(b-e) have been proven to occur in models containing the lateral matrix.

**2.4. Classical Nucleation Theory (CNT) calculations**

For $Mg_2Ca$ homogeneous nucleation, we simplify Eq. (6) as:

$$\Delta G_{\mathrm{Mg_2Ca}}=n\Delta\mu_{\mathrm{Ca}}+\frac{2nV_0}{t}\gamma_{\mathrm{b}}+6t\sqrt{\frac{2nV_0}{3\sqrt{3}t}}\gamma_{\mathrm{r}}+\Delta G_{\mathrm{elastic}} \quad (1)$$

Here $n$ is the number of Ca atoms in the nucleus, $\Delta\mu_{Ca}$ is the chemical potential difference between Ca in the $Mg_2Ca$ phase and the solid solution state of the Mg matrix, and $V_0$ represents the volume of a single $Mg_2Ca$ formula.

Since there exists little thermodynamic data about $Mg_2Ca$ phase, we herein substitute $V\Delta G_{\mathrm{chem}}$ term in Eq. (6) with $n\Delta\mu_{\mathrm{Ca}}$ term[4,8], where chemical potential $\mu_{\mathrm{Ca}}$ in $Mg_2Ca$ condition ($\mu_{\mathrm{Ca\text{-}Mg_2Ca}}$) and $\mu_{\mathrm{Ca}}$ in solid-solution condition ( $\mu_{\mathrm{Ca\text{-}solid}}$ ) are calculated by the following relations: $\mu_{\mathrm{Mg}}=E_{\mathrm{hcp\text{-}Mg}}$, $2\mu_{\mathrm{Mg}}+\mu_{\mathrm{Ca\text{-}Mg_2Ca}}=E_{\mathrm{Mg_2Ca}}$, $\mu_{\mathrm{Ca\text{-}solid}}=E_{\mathrm{Mg_{47}Ca}}\text{-}47E_{\mathrm{Mg}}$ , $\Delta\mu_{\mathrm{Ca}}=\mu_{\mathrm{Ca\text{-}Mg_2Ca}}\text{-}\mu_{\mathrm{Ca\text{-}solid}}$ . Although the entropic contributions (vibrational or configurational) are not incorporated in this reduced method, earlier comparable works [4,8,14] have demonstrated that this treatment is capable of producing acceptable results.

For interfacial energies of coherent broad interface $\gamma_{\mathrm{b}}$ and the rim interface $\gamma_{\mathrm{r}}$, we construct supercells of $Mg_2Ca$ and $\alpha$-Mg to calculate these two quantities. The energy of formation per atom relative to the energies of $\alpha$-Mg and $Mg_2Ca$ phases can be written as follows[36,37]:

$$\Delta E_f=\delta E_{\mathrm{cs}}+\frac{S_{\mathrm{b}}\gamma_{\mathrm{b}}}{N} \quad (2)$$

$$\Delta E_f=\delta E_{\mathrm{cs}}+\frac{S_{\mathrm{b}}\gamma_{\mathrm{b}}+S_{\mathrm{r}}\gamma_{\mathrm{r}}}{N} \quad (3)$$

Where $\Delta E_f$ is the energy of supercell formation (eV/atom) relative to the bulk energies of $\alpha$-Mg and $Mg_2Ca$ phase. $\delta E_{\mathrm{cs}}$ is the coherency strain per atom caused by the lattice mismatch between $\alpha$-Mg and $Mg_2Ca$ phases. $N$ is the total number of atoms in the supercells, and $\gamma_{\mathrm{b(r)}}$ and $S_{\mathrm{b(r)}}$ are the interfacial energy and area, respectively. The interfacial energy can be obtained from the slope of $\Delta E_f$ (eV/atom) *vs*. $1/N$ as shown in Fig. S11. Here, to rigorously assess the $\gamma_{\mathrm{r}}$ value, we first used Eq. (2) to calculate the $\gamma_{\mathrm{b}}$, and this value is then taken into Eq. (3) to obtain the $\gamma_{\mathrm{r}}$. Our obtained $\gamma_{\mathrm{b}}$ and $\gamma_{\mathrm{r}}$ values are 0.13 J/m$^2$ and 0.41 J/m$^2$, respectively.

With the assumption of homogeneous and isotropic elasticity, for a penny-shaped nucleus with a radius $R$ and an aspect ratio ($a$=2$R$/$t$), $\Delta G_{\text{elastic}}$ can be analytically calculated from the Eshelby's solution[38]:

$$\Delta G_{\text{elastic}}=\mu\Big[ \left({\varepsilon_{11}^{*}}^{2}+{\varepsilon_{22}^{*}}^{2}\right)\left(\frac{v}{1\text{-}v}-\frac{13}{32(1\text{-}v)}\frac{\pi}{a}+1\right)+{\varepsilon_{33}^{*}}^{2}\frac{1}{4(1\text{-}v)}\frac{\pi}{a}+2\varepsilon_{11}^{*}\varepsilon_{22}^{*}\left(\frac{v}{1\text{-}v}-\frac{16v\text{-}1}{32(1\text{-}v)}\frac{\pi}{a}\right)+\left(2\varepsilon_{11}^{*}\varepsilon_{33}^{*}+2\varepsilon_{22}^{*}\varepsilon_{33}^{*}\right)\frac{2v+1}{8(1\text{-}v)}\frac{\pi}{a}+\left(2{\varepsilon_{23}^{*}}^{2}+{\varepsilon_{31}^{*}}^{2}\right)\frac{2\text{-}v}{4(1\text{-}v)}\frac{\pi}{a}+2{\varepsilon_{12}^{*}}^{2}\left(1-\frac{7\text{-}8v}{16(1\text{-}v)}\frac{\pi}{a}\right)\Big] \quad (4)$$

Where $\varepsilon_{ij}^{*}$ are components of the stress-free strain tensor associated with lattice mismatch, $\mu$ the shear modulus and $v$ the Poisson ratio. The Poisson's ratio for Mg of $v$=0.29 was used. Because the shear modulus $\mu$ of Mg and $Mg_2Ca$ phase have similar values, i.e., $\mu_{\text{Mg}}$=18 GPa and $\mu_{\text{Mg}_2\text{Ca}}$=19 GPa [39], we chose 18 GPa to perform calculations. The earlier work[36] shows this estimation method usually overestimates the contribution of $\Delta G_{\text{elastic}}$ owing to elastic relaxation effects during precipitation.

Different from Ca serves as large atoms in the $Mg_2Ca$ phase, Zn plays small atoms in the $MgZn_2$ phase. Hence, the total energy change associated with $MgZn_2$ homogeneous nucleation can be written as:

$$\Delta G_{\text{MgZn}_2}=n\Delta\mu_{\text{Zn}}+\frac{nV_0}{t}\gamma_{\text{b}}+6t\sqrt{\frac{nV_0}{3\sqrt{3}t}}\gamma_{\text{r}}+\Delta G_{\text{elastic}} \quad (5)$$

where $n$ is the number of Zn atoms in the nucleus, $\Delta\mu_{\text{Zn}}$ is the chemical potential difference (i.e., $\Delta\mu_{\text{Zn}}=\mu_{\text{Zn-MgZn}_2}\text{-}\mu_{\text{Zn-solid}}$) between Zn in the $MgZn_2$ phase and the solid solution state of the Mg matrix, and $V_0$ represents the volume of a single $MgZn_2$ formula. The $\Delta\mu_{\text{Zn}}$, interfacial energies $\gamma_{\text{b}}$ and $\gamma_{\text{r}}$ and $\Delta G_{\text{elastic}}$ contributions are all calculated using the same way as introduced above. As shown in Fig. S11(b), our calculated $\gamma_{\text{b}}$ and $\gamma_{\text{r}}$ values are 0.21 J/m$^2$ and 0.40 J/m$^2$, respectively. Although the shear modulus of the $MgZn_2$ phase is 27 GPa, our tests and other previous works [14,36] have proved that this value has a negligible impact on the final results, thereby we still chose $\mu_{\text{Mg}}$=18 GPa to calculate $\Delta G_{\text{elastic}}$ contributions.

**2.5 Phonon dispersion curves**

In the calculation of phonon dispersion (Fig. 4), the harmonic interatomic force constants were obtained by density functional perturbation theory (DFPT) using the supercell approach, which calculated the dynamical matrix through the linear response of electron density[40]. The Natarajan et al.[27] model and 2D-infinite model, respectively, employed the $3\times3\times2$ and $3\times2\times1$ supercells. The phonon dispersion was obtained by using the *phonopy* code[41] with the harmonic interatomic force constants as input.

## 3. Results

### 3.1. A new definition for TCP structures

Typically, coherent precipitates formed in situ within a matrix have similar lattice configurations to those of the matrix on specific atomic planes (i.e., habit planes)[42], which means that only slight local lattice rearrangements are required to achieve lower interfacial and strain energies, thereby facilitating nucleation. Our results below indicate that the fundamental reason for the coherent precipitation of TCP phases within the hcp-Mg matrix lies in the crystallographic similarity between the TCP and hcp structures. Thus, to enhance comprehension of this hcp→TCP structural transformation, we initially introduce the basic features of TCP structures.

Among the various TCP members, to our knowledge, only Laves and Laves-like precipitates [43,44] have been explicitly found in the hcp-Mg matrix, including $Mg_2Ca$[11,45], $Al_2Ca$[20,21], $Al_2Gd$[46], $MgZn_2$[11,47], $\gamma''$ phases[19,43,44] and their numerous derivatives[11,22]. These TCP plates often display plate-shaped morphologies and maintain fixed orientation relationships (OR: $[11\bar{2}0]_{C14/C36}//[10\bar{1}0]_{Mg}$, $(0001)_{C14/C36}//(0001)_{Mg}$, or $[01\bar{1}]_{C15}//[10\bar{1}0]_{Mg}$, $(111)_{C15}//(0001)_{Mg}$) with respect to the matrix. Other common TCP members such as the the $\sigma$, $\mu$, P, and R phases often precipitate in fcc-based superalloys[48–50] or bcc-based stainless steels[49] but have not been observed in an hcp-based matrix. Therefore, this work solely focuses on the formation pathways of Laves and Laves-like (including $CaCu_5$-, $Ce_2Ni_7$-, $Y_2Ni_7$- and $NbBe_3$-type prototypes) phases, with Laves phases being discussed in the main text and Laves-like phases in the Supplementary Materials.

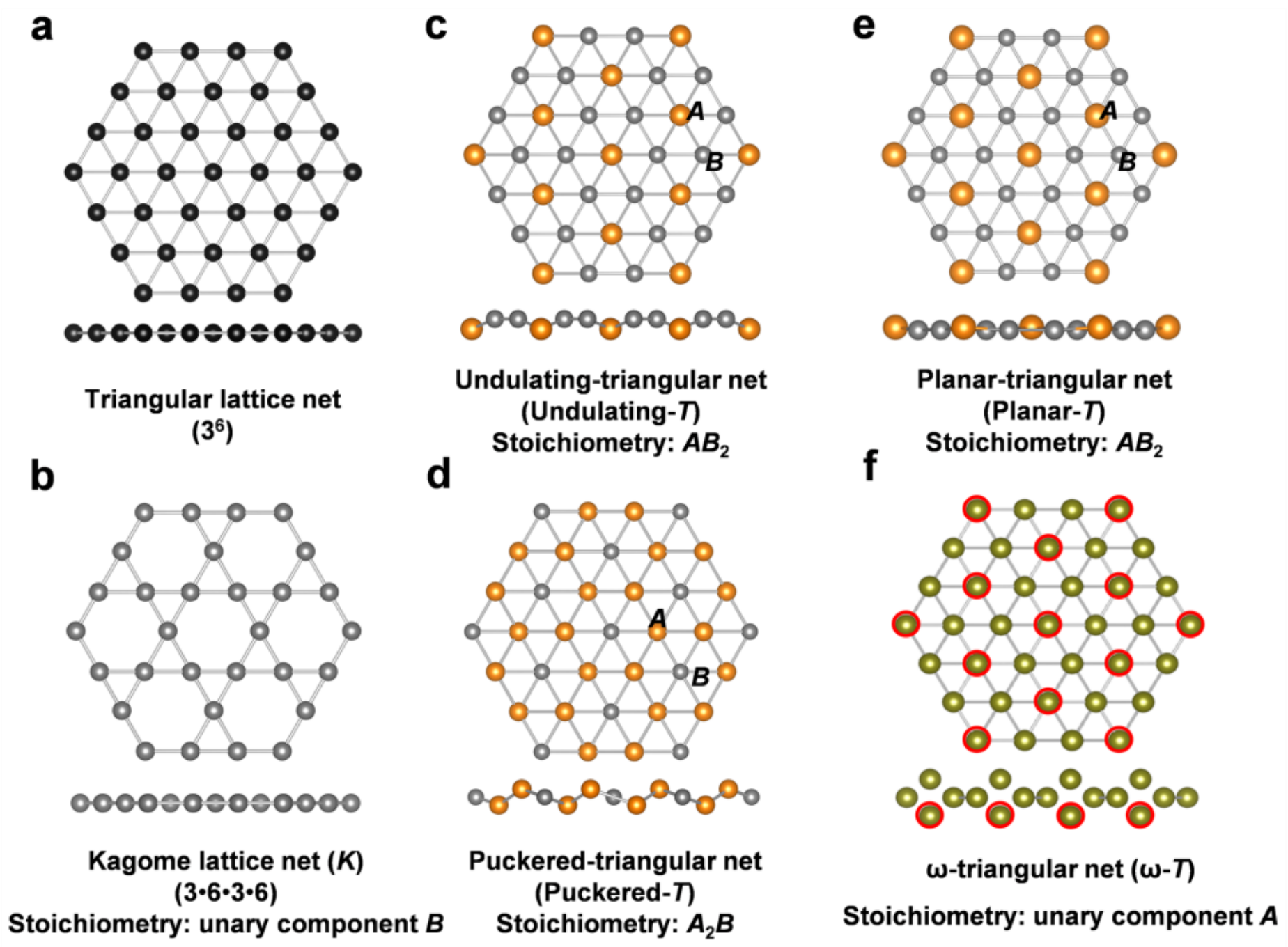

**Fig. 1.** Atomic packing patterns of triangular and kagome lattice nets in the TCP phases. (a) The packing pattern of a regular triangular lattice net that constructs pure metals. (b) The atomic configuration of a normal kagome lattice net in the TCP phases, which is made up of small atoms *B*. (c-e) The atomic packing patterns of the undulating triangular net ($AB_2$), puckered triangular net ($A_2B$), and planar triangular net ($AB_2$). Large-yellow balls and small-grey balls represent the *A*-type (large atoms) and *B*-type (small atoms) elements, respectively. (f) The atomic structure of the ω triangular net. This ω-*T* often appears in the $\mu$ and $Zr_4Al_3$-type phases, we thus do not examine this net in the current work owing to its correlation with the additional interstitial atoms (indicated by red-edged green balls). Each panel's lower side patterns display horizontal views of the corresponding upper side patterns.

The TCP structures are typically described as lamellar compounds consisting of alternating primary and secondary layers[17,18]. The Laves and Laves-like phases, for instance, are composed of triangular (*T*) and kagome (*K*) layers in a pattern of ···*TKTKT*···. In Fig. 1, we present the triangular (*T*) and kagome (*K*) lattice nets that occur in the Laves and Laves-like phases. To aid in differentiation, we categorize the triangular lattice nets into four different sub-triangular lattice nets based on their projected side views. It is worth noting that the *K* lattice net typically keeps a planar morphology, and its atomic density per unit cell ($\rho$) equals that of *T* nets (i.e., $\rho_K = \rho_T = 3$). Furthermore, we emphasize that the undulating-*T* lattice net does not occur within TCP structures but rather exists as a coherent TCP/matrix broad interface adjacent to the {0001}hcp basal planes, as will be demonstrated later.

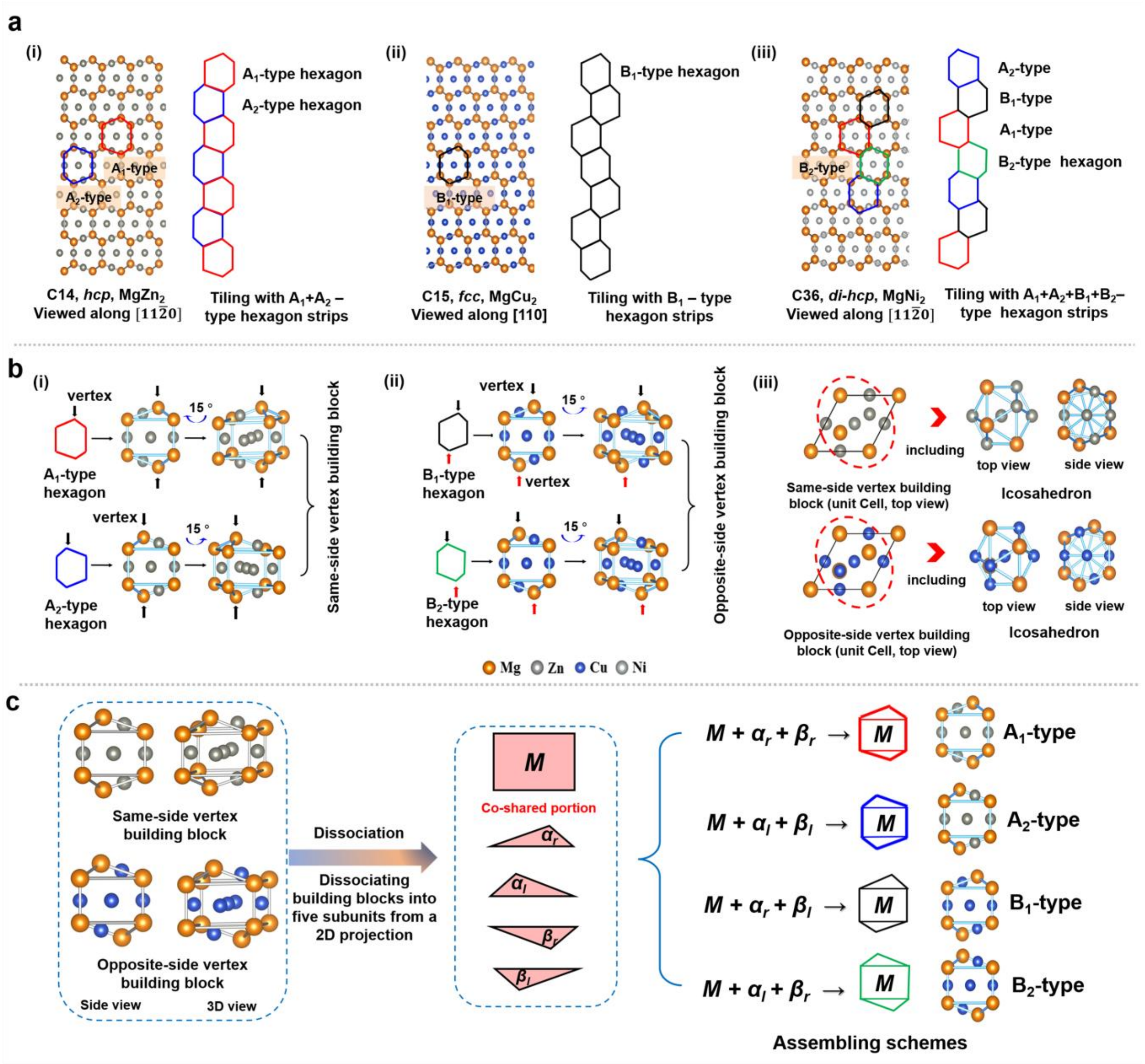

**Fig. 2.** Schematic illustrations of a new definition for TCP structures. (a) Three polytypes of Laves phases are assembled using abstracted building blocks. (i-iii) $MgZn_2$ C14 phase, $MgCu_2$ C15 phase, and $MgNi_2$ C36 phase can be viewed as stacking by the four kinds of hexagons in 2D projections, where these hexagons are shown as red-edged $A_1$-type, blue-edged $A_2$-type, black-edged $B_1$-type, and green-edged $B_2$-type hexagons, respectively. And the yellow, dark-grey, dark-blue, and light-grey balls indicate the Mg, Zn, Cu, and Ni atoms, respectively. (b) (i-ii) The actual crystallographic structure of four kinds of hexagons, where $A_1$ and $A_2$-type hexagons belong to the same-side vertex (marked as black arrows) building block and the other two belong to the opposite-side vertex building block (labeled as black and red arrows). (iii) Atomic structure of icosahedron in the two types of building blockings. (c) Dissociative five subunits of building blocks in the Laves phases. The combinational scheme for building four types of building blocks is shown on the right side.

Indeed, this lamellar definition provides clear structural features of the TCP phases, but it is difficult to correlate these features with the matrix upon initial nucleation. We thus propose a new crystallographic definition for TCP structures, as illustrated in Fig. 2a. In two-dimensional (2D) projections, the three basic Laves polytypes, C14, C15, and C36, can be represented as a tilling made up of four types of non-equilateral hexagons (designated as $A_1$, $A_2$, $B_1$, and $B_2$-type hexagons). These hexagons correspond to the basic structural unit of the Laves phases. In their 3D formations, Fig. 2b (i-ii) indicates that these four hexagons are essentially two types of polyhedra, with their vertices on the same side (black arrows) or the opposite side (red arrows). The $A_1$- and $A_2$-type belong to the same-side vertex building block, while the $B_1$- and $B_2$-type hexagons belong to the opposite-side vertex building block. According to this new definition, all Laves structures can be constructed entirely from these two types of building blocks, producing four distinct hexagons resulting from various 2D projections. Additionally, both building blocks contain an icosahedron (red-dotted oval) within their structure (Fig. 2b(iii)). This icosahedral definition has also been commonly used in earlier investigations [44,51].

In 2D projections (Fig. 2c), a careful examination of the two types of building blocks reveals a general geometric feature: each hexagon consists of a co-shared rectangle and two triangular parts. Furthermore, these two types of building blocks can be further dissociated into five subunits: one rectangle $M$ subunit and four triangles ($\alpha_r$, $\alpha_l$, $\beta_r$, and $\beta_l$) subunits. With these five subunits, the original $A_1$, $A_2$-type, and $B_1$, $B_2$-type hexagons can be reconstructed using four different combinations. Similarly, the Laves-like phases can also be described using these dissociative subunits, as indicated in Supplementary texts-II. Importantly, this new structural definition identifies a co-shared portion among distinct TCP configurations. This enables us to determine the basic structural transformation unit (BSTU) of hcp→TCP structural transformations and describe the entire process of structural evolution in the subsequent section.

### 3.2. Structural evolution pathway for Laves phases

In the conventional view, the description of precipitate formation should start from the nucleation stage. A mark for the end of nucleation stage is the formation of the critical nucleus that can stably exist and grow instead of dissipating. In contrast to isostructural transformations with negligible structural variations, the formation of complex TCP structures within a hcp-Mg matrix should involve noticeable structural changes, and these structural transformations lead to a decrease in system energy. Hence, structural transformation has already occurred before the nucleus evolved into a critical nucleus. The results below indicate that the TCP phase appears with multilayer TCP structure in the hcp-Mg matrix each time, rather than the traditional layer-by-layer perspective. Consequently, we first introduce an evolution pathway of TCP structures in the hcp-matrix, which outlines the atomic structure evolution of TCP plates within the hcp-matrix, and further evaluate the size of the critical nucleus using this pathway in section 4.2.

Here, we begin by using the $Mg_2Ca$ phase as an illustrative example to demonstrate how this Laves phase with simple composition (i.e., containing Mg) progressively evolved from the hcp-Mg matrix through a 3-layer unstable hcp-ordering structure. Then, the scope is extended to other TCP phases without Mg ( such as the $Al_2Ca$ phase, etc.). At low temperatures, the $Mg_2Ca$ Laves phase is an equilibrium phase in the Mg-Ca binary alloy, having a C14 structure [39,45]. Inspired by the instability of specific hcp-orderings proposed by Natarajan [27] et al., some solute clusters with certain structures may induce structural transformations. Nonetheless, the genuine structure of these specific orderings within the hcp-matrix remains unclear due to their neglect of the matrix's role (for details see Supplementary texts-I). Thus, we adopt the model containing the matrix portion to explore the true structure of unstable solute orderings.

Although ab initio calculations are highly accurate, they are also expensive and time-consuming, making it impractical to directly search for unstable hcp-orderings using random sampling methods in large-scale models. However, the proposed new TCP definition can help capture these unstable hcp-orderings using finite samplings. Fig.2(c) shows that the *M*-type structure is the co-shared portion in both types of building blocks. Therefore, we first employ the 270-atom model (Method) in Fig. 3(a) to build (i.e., sampling) an $M_R$-type Ca hcp-ordering possessing a unit-cell size in the hcp-Mg lattice, which shares the same relative arrangement of Ca atoms as that *M*-type structure. After a structural relaxation, the noticeable lattice distortion ($\Delta d \approx 0.33$Å, $\Delta\theta = 36.19°$) indicates that this unit-cell $M_R$-type Ca ordering seems unstable within the hcp-Mg lattice. Then in Fig. 5(b), if the size of this $M_R$-type Ca hcp-ordering is further expanded along the $\{0001\}_{Mg}$ basal planes (which has been demonstrated to be an energetically favorable process in Supplementary text-IV), this larger-sized ordering will gradually transform into a *M*-type structure upon relaxation.

To exclude the influence of periodic boundary on the results, we repeat this process in Fig. A1 using a 450-atom model (Methods) containing a larger lateral matrix. The results still indicate that this $M_R$-type hcp-ordering is unstable and will transform into an *M*-type structure progressively. Besides, Fig. A2 also illustrates that the presence of vacancies in $M_R$-type ordering does not affect this structural transformation. Therefore, although the atomic migration at the microscale is stochastic and the solute clusters generated by compositional fluctuations are diverse, only when the Ca-clusters evolve to 3-layer $M_R$-type ordering at a specific moment can these Ca-clusters spontaneously collapse into an *M*-type structure. Additionally, given the absence of accurate interatomic potentials [4], it is unfeasible to utilize Molecular Dynamics or Kinetic Monte Carlo techniques to explore the long-term dynamic evolution of solute clusters. However, in ab initio calculations, the finite samplings through capturing unstable hcp-orderings that lead to structural transformation at several specific states can still provide key node details about long-term structural evolution. This is also the basis of using ab initio calculations to explore the structural evolution of TCP structures within an hcp matrix.

During structural transformation, Fig. 3(b) shows that collective displacement $\xi_x$ (labeled as black arrows) transforms the original $T$ net in the central plane of $M_R$-type ordering into a $K$ net (θ=180°), while dissociative displacement $\xi_y$ splits the outermost two $T$ nets of $M_R$-type ordering into two undulating-$T$ nets. To avoid overcrowding, the $T\rightarrow K$ transition converts previous clusters centered at Ca atoms into hexagonal openings. Overall, the $\xi_x$ and $\xi_y$ displacements together complete the structural transformation. The $M$-type structure can be recognized as the $CaCu_5$-*split* structure owing to its dissociative outermost two $T$ lattice nets. Consequently, the formation of an $M$-type structure signifies the presence of a TCP structure within an hcp lattice.

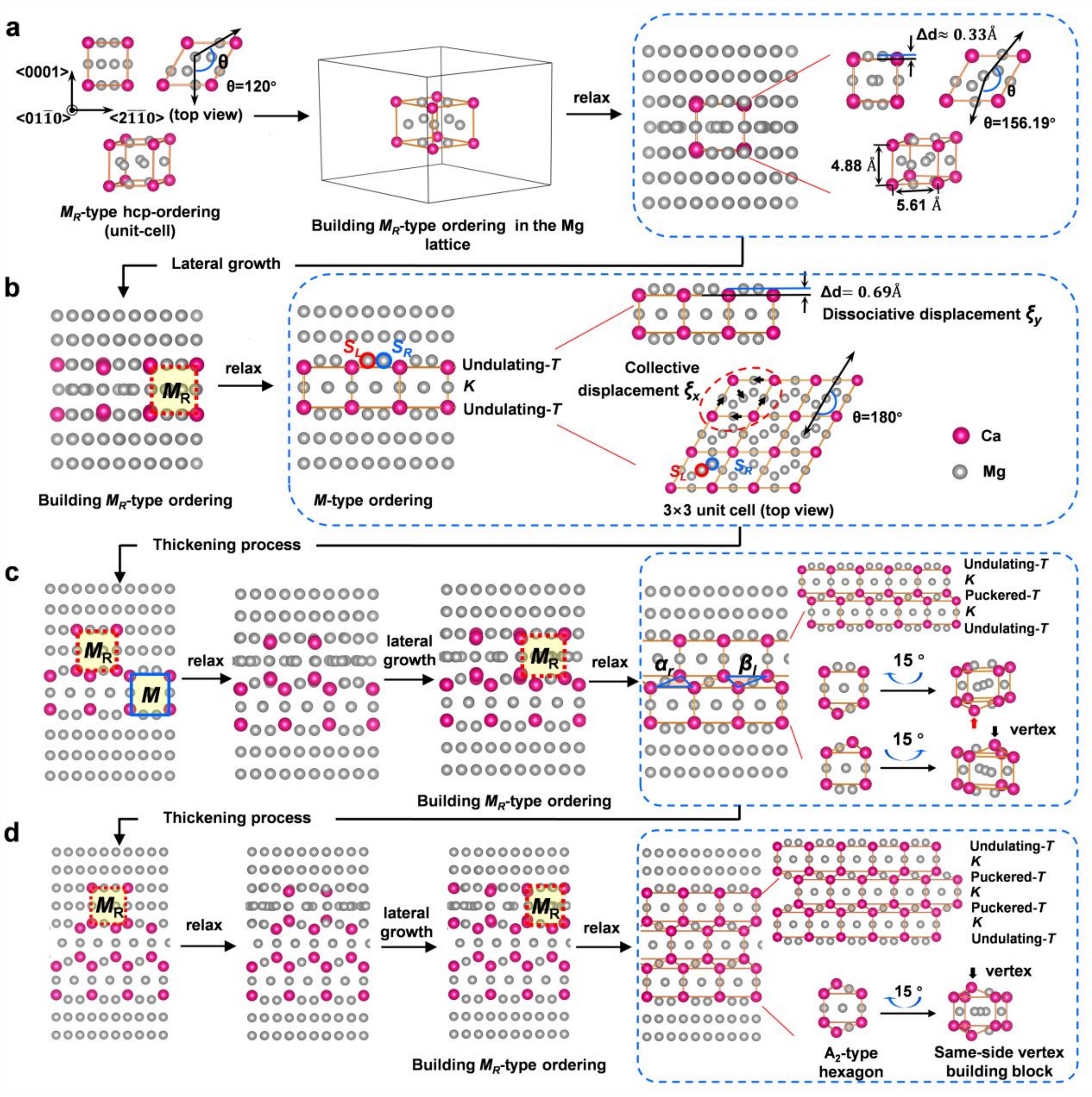

**Fig. 3.** The structural evolution pathway for $Mg_2Ca$ Laves phase in the hcp-Mg matrix. (a-b) A nucleation process for the *M*-type subunit of the $Mg_2Ca$ phase, with 3-layer $M_R$-type orderings (indicated by a red-dotted rectangle) as the basic structural transformation unit. The Mg and Ca atoms are denoted by grey and dark-pink balls, respectively. And the $S_L$ and $S_R$ atomic sites in (b) are enclosed with red and dark-blue circles, respectively. (c-d) The thickening process for the $Mg_2Ca$ phase. Through building $M_R$-type orderings at the $Mg_2Ca$/matrix interface, $B_1$-type building block forms in the matrix. The accurate structural analysis and the change in structure that results from transformation are shown on the right side of each panel.

In particular, despite that experimentally observed $Mg_2Ca$ and other Laves precipitates generally exhibit substantially thickened (Fig. A3), the direct experimental evidence for this 3-layer *M*-type structure can still be found in the $\gamma''$ phase of Mg-RE-Zn series alloys[43,44,52]. As the thinnest TCP structure ever detected in Mg alloys, the metastable $\gamma''$ phase often keeps an *M*-type structure throughout the aging treatment, thus confirming the existence of the 3-layer *M*-type structure. Additionally, using the phonon dispersion curves given in Fig. 4, we stress that the spontaneous transformation of unstable hcp-orderings into TCP structures during structural relaxation is caused by internal-coordinate instability[52], rather than the lattice dynamical instability proposed by Natarajan et al.[27,53], due to this structural transformation occurs locally within the matrix and Fig. 4 does not show any imaginary frequencies.

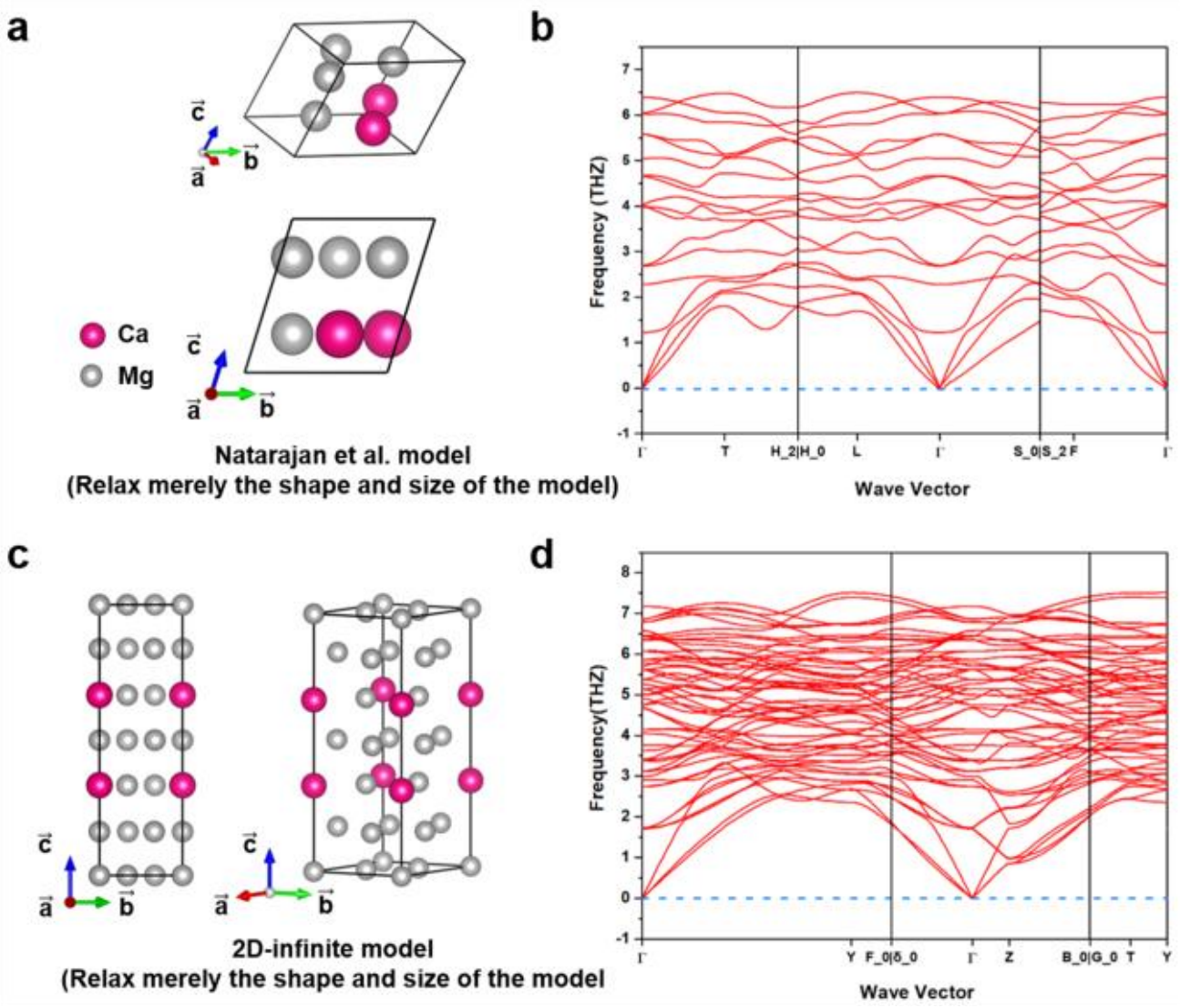

**Fig. 4.** Calculated phonon dispersion curves for the Natarajan et al. [27] model (a-b) and the 2D-infinite model (c-d). To restrict the freedom of internal atomic coordinates, we simply relax the model's size and shape during the relaxation process. These two phonon dispersion curves do not contain any imaginary frequencies, proving that the hcp-to-TCP structural transformations do not associate with the dynamical instabilities while belonging to the internal-coordinates instabilities.

Reexamining the *M*-type structure (Fig. 3(b)), the periodicity along the $[0001]_{Mg}$ direction of this structure has been destroyed by the outward Mg atoms in the outermost two undulating-*T* nets. And these two undulating-$T$ lattice nets maintain almost full coherency with the matrix. How to thicken such a coherent interface becomes a key problem for subsequent growth processes. By comparing the differences in crystal structure between the $M$-type structure and the two building blocks of Laves phases (Fig. 2(b)), we show that the thickening process is associated with modifying the atomic structure of the undulating-$T$ lattice nets. Two routes for breaking the constraint of undulating-$T$ lattice nets are proposed in this study, one specifically for the Laves phase and the other for Laves-like phases.

Fig. 3(b) shows the undulating-$T$ nets have one-third of their atomic sites occupied by Ca atoms, while the remaining atomic sites can be divided into two categories: $S_R$ (dark-blue circle) and $S_L$ (red circle) atomic sites. Although $S_R$ and $S_L$ atomic sites are symmetrical, we show below that different polytypes of Laves phases are closely related to these two atomic sites, which prompts us to make this distinction. Afterward, in Fig. 3(c), we further constructed $M_R$-type Ca orderings on the $S_R$ atomic sites. An evident 5-layer thickened structure can be formed by extending these orderings laterally. Similarly, this lateral growth linked with long-range solute diffusion has also been proved to be an energetically preferred process in Supplementary texts-IV. During thickening, the undulating-$T$→puckered-$T$ transformation produces $\alpha_r$ and $\beta_l$ subunits. And the whole thickening process is still completed by the tiny $\xi_x$ and $\xi_y$ displacements gradually. A slight difference from that of forming an $M$-type structure is that the undulating-$T$ lattice net consisting of two sublayers further splits into a three-sublayer puckered-$T$ lattice net.

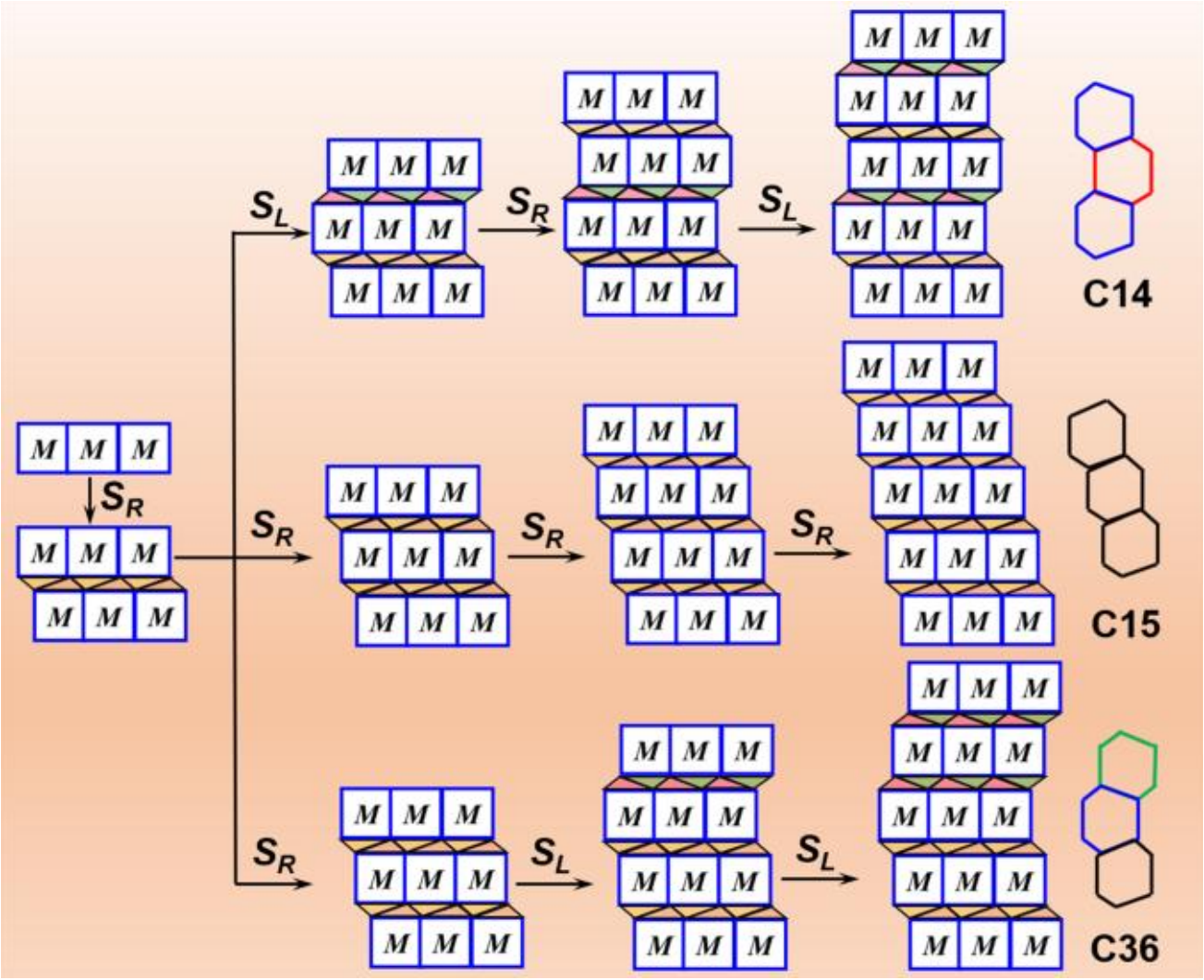

**Fig. 5.** Formation pathways of three polymorphic Laves phases in hcp matrix. $S_R$ and $S_L$ indicate the atomic sites shown in Fig. 3(b). The various types of basic building blocks can be created by building $M_R$-type ordering on the $S_R$ or $S_L$ atomic site.

In this way, Fig. 3(d) shows the 5-layer structure can be further transformed into a 7-layer structure by establishing $M_R$-type orderings on the undulating-$T$ lattice net, with the same-side vertex building blocks (i.e., $A_2$-type hexagon) form in this 7-layer configuration. By employing this scheme in Fig.5, the different types of Laves phases can be formed through the establishment of $M_R$-type orderings on either the $S_R$ or $S_L$ atomic sites. Furthermore, these Laves phases may also undergo mutual transformations via a synchro-shear mode, as a typical example of C14→C15 illustrated [54]. However, given the complex-to-complex nature of these mutual transformations, we do not discuss this topic here. Overall, as depicted in Fig. 6, the whole structural evolution in the formation of Laves phases is fully accomplished by these 3-layer $M_R$-type hcp-orderings via shuffle-based displacements. These inherently unstable 3-layer $M_R$-type hcp-orderings can be viewed as the BSTU that governs the entire structural transformation process.

Additionally, the hcp→TCP transition is a conservative transformation without requiring extra atoms in that the same atomic density per unit area of $T$ and $K$ lattice nets (i.e., $\rho_K = \rho_T = 3$). Unlike the traditional viewpoint [7,13] of dislocation-mediated ledge lateral migration, as famous cases found in Al-Ag [55] and Ti-Al [56] systems, the lengthening of Laves plates relies on forming $M_R$-type orderings along their minor facets (i.e., the growth ledge lacks dislocation character, as shown in Fig. 9(a)) continuously. This is mainly attributed to the $T \rightarrow K$ transition cannot be simply achieved through the glide of specific dislocations but rather collective displacement $\xi_x$. Moreover, the most important ledge nucleation is dependent on modifying the interface structure via forming $M_R$-type orderings on their original undulating-$T$ lattice nets. Meanwhile, in Supplementary texts-IV, it is demonstrated that the lengthening process is preferable over the thickening process due to the distinctly different abilities of the rim interface in attracting solutes relative to the broad interface. As a result, the TCP precipitates readily form with a plate-like morphology.

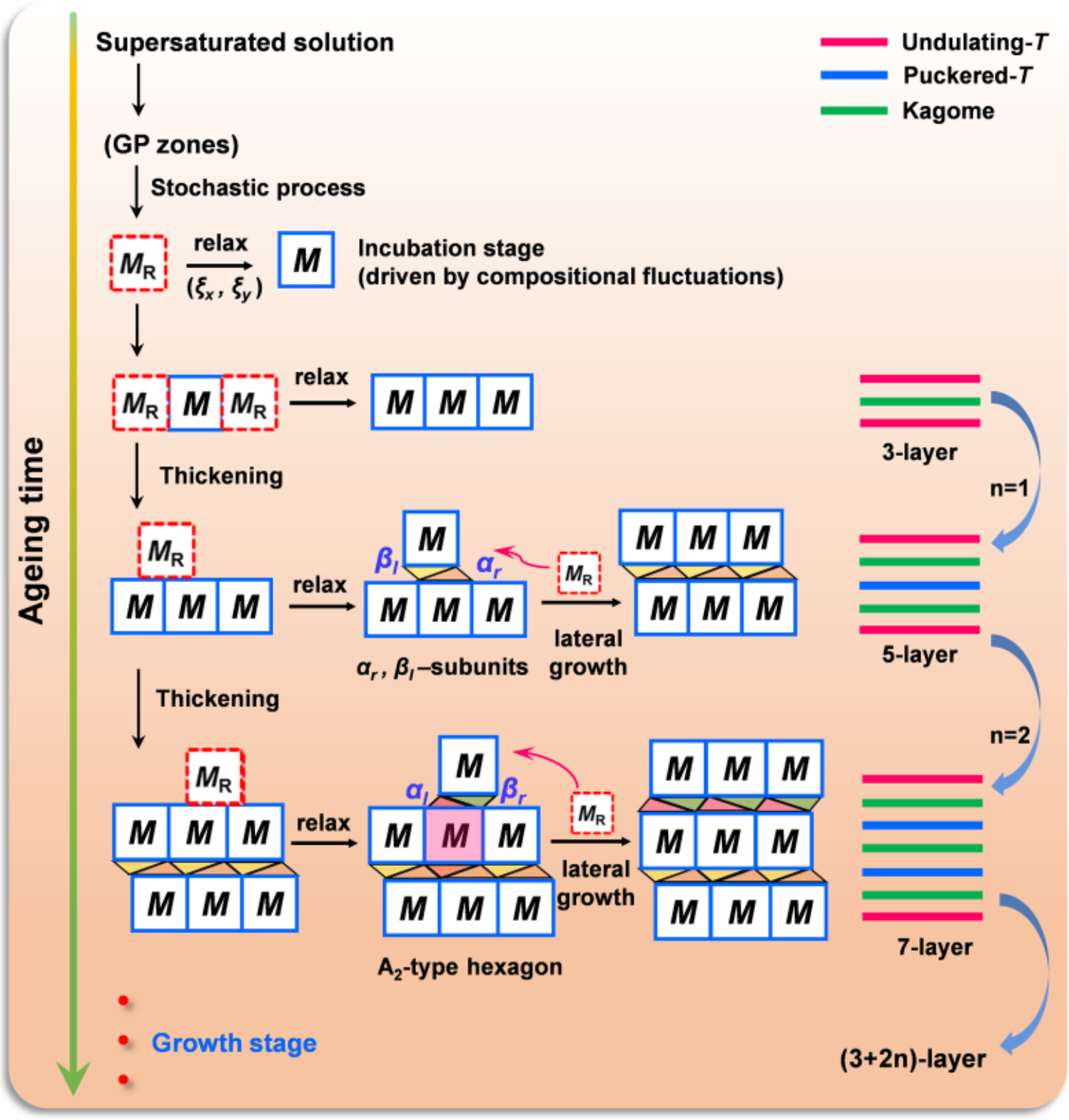

**Fig. 6.** A schematic of the structural evolution pathway for the Laves phase within the hcp matrix. The red-dotted and dark-blue rectangles denote the $M_R$-type and $M$-type orderings, respectively. The n represents the number of thickening steps of Laves phase.

Regarding other Laves and Laves-like phases (Supplementary texts-III), their formation pathways are basically the same as that of the $Mg_2Ca$ phase: the entire structural evolution is still accomplished by establishing $M_R$-type hcp-orderings. Somewhat differently, to induce a noticeable structural transformation, for example, the occurrence of structural transformation for the $MgZn_2$ phase (Fig. A4) necessitates the presence of Zn atoms around the Mg atoms (i.e., larger-sized $M_R$-type Zn orderings). This difference primarily relates to the kinetic conditions of structural transformations that will be discussed below. Moreover, Mg-free TCP phases exhibit more nonclassical nucleation characteristics. For instance, as shown in Fig. A5(a-c), the nucleation of the $Al_2Ca$ phase does not require full occupancy of smaller atomic sites by Al atoms. Instead, it primarily relies on the arrangement of large Ca atoms. Such a form of nucleation with a non-uniform composition within the nucleus is a typical nonclassical nucleation process. Furthermore, even if Ca atoms occupy the anti-sites (Fig. A5(d-f)), the formation of the $Mg_2Ca$ phase can still occur through this nucleation mechanism, suggesting that nonclassical nucleation is a fundamental feature in the nucleation of TCP precipitates irrespective of the Mg-free or Mg-containing product phases. Additionally, the presence of polytypes and off-stoichiometry in TCP precipitates are also associated with the nonclassical nucleation behavior, as demonstrated in Supplementary text-V.

## 4. Discussions

### 4.1. BSTU in hcp-to-TCP structural transformations

Theoretically, the hcp-matrix, as the parent crystal, offers natural advantages in the formation of Laves and Laves-like phases. This is because the similar stacking sequence between the hcp (*ABABA*) lattice and the TCP structures (*TKTKT*) only requires the close-packed $\{0001\}_{hcp}$ basal planes to undergo modest shuffle-based displacements to complete the structural transformation. As the thinnest structure that forms a TCP structure in the hcp-matrix, the 3-layer $M_R$-type hcp-ordering is recognized as the BSTU of the whole structural evolution in the formation of various Laves and Laves-like phases, as shown in Fig. 7. Meanwhile, considering the nonclassical nucleation nature of hcp→TCP transformations, the atomic structure of unstable $M_R$-type hcp-ordering is not fixed. More importantly, we emphasize the role of the BSTU in a simple-to-complex structural transformation, because this BSTU governs the entire structural evolution of the product phase within the parent phase.

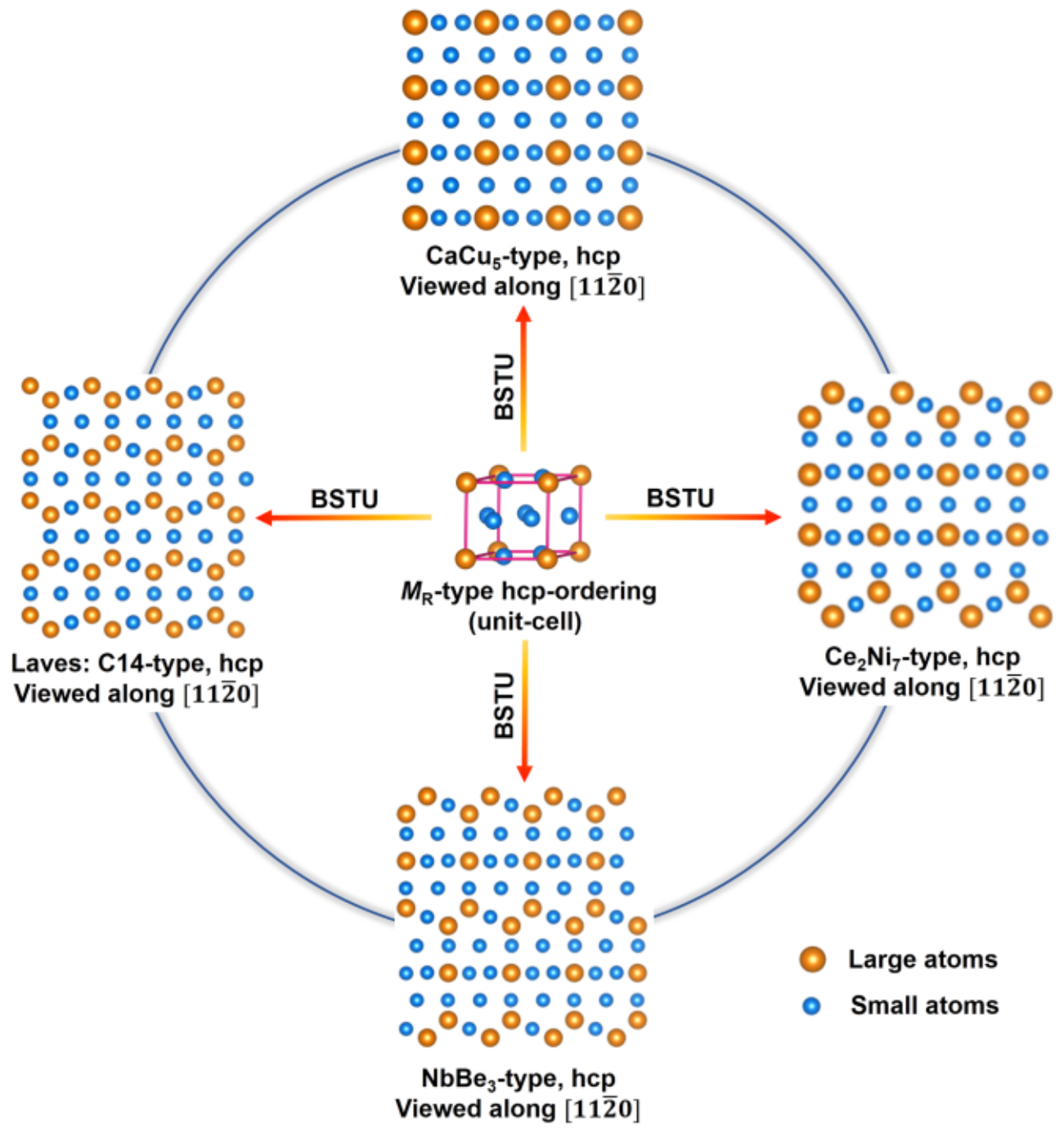

**Fig.7.** A schematic illustration of the relationship between the 3-layer $M_R$-type hcp-ordering and diverse TCP phases. The yellow and light-blue spheres represent large and small atoms, respectively.

However, why are the TCP structures formed with a 3-layer as the basic unit within the hcp matrix? In Fig. 6, to address this problem, we map the energy($(E(\xi_x, \xi_y)$- $E(1, 1))$ landscape of structural transformation relative to normalized collective displacements $\xi_x$ and dissociative displacements $\xi_y$, using the energy $E(1, 1)$ of final relaxed structures as the reference (Section 2.4). In a structural transformation, the system always follows a minimum energy path (MEP) to complete the process. During stage i (Fig. 8(b-c)), both the Mg-Gd and Mg-Ca systems indicate that synchronous displacements of $\xi_x$ and $\xi_y$ result in the steepest descent of system energy along the black line (MEP). This means that $\xi_x$ and $\xi_y$ displacements are interdependent in hcp→TCP transformations; if either $\xi_x$ or $\xi_y$ is constrained, the structural transformation cannot occur spontaneously due to a significant energy barrier. This conclusion also applies to the thickening processes of $Mg_2Ca$ and $Ce_2Ni_7$ phases (Figs. 8(d-e)).

Therefore, collective displacement must be accompanied by the dissociative displacement of the two *T* lattice nets adjacent to it, and *vice versa*. This demonstrates that the 3-layer structure is an essential prerequisite (i.e., a kinetic condition that triggers the structural transformation) for the initial nucleation and the two basal planes act as the minimum thickening unit at each thickening step. Here, we formulate this property as $N$=3+2$n$, where $N$ represents the total number of layers in TCP precipitates and $n$ specifies the number of thickening steps. As indicated in Fig. A3, the validity of this odd-layer property of TCP precipitates is supported by various experimental characterizations [21,22,44,57].

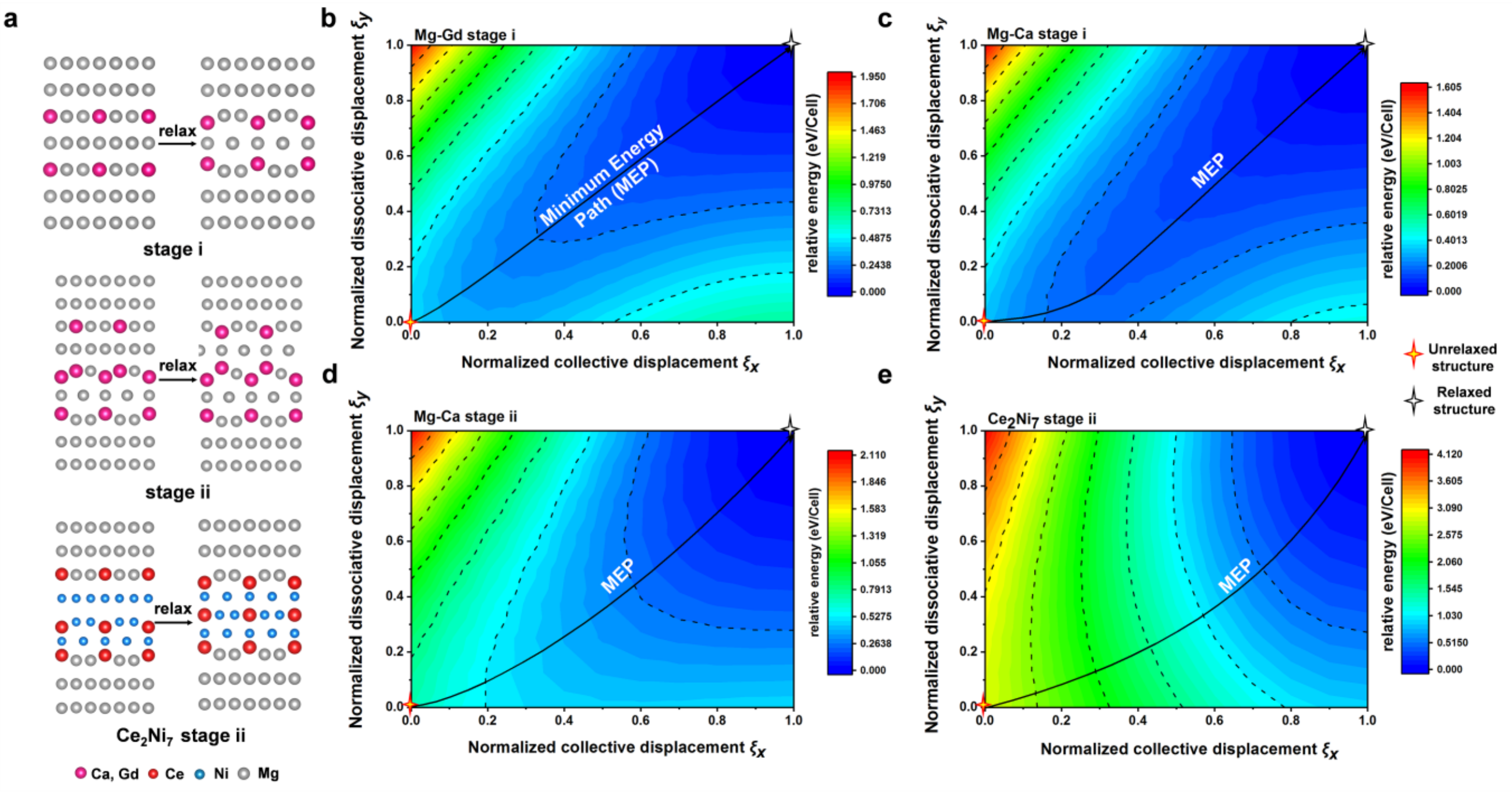

**Fig. 8.** Energy landscape of structural transformations connecting the initial unrelaxed structures to the final relaxed structures. (a**)** The atomic configurations of three calculated stages. The initial unrelaxed and final relaxed structures are shown on the left and right sides of each panel. Stage i corresponds to the initial nucleation stage of Mg-Ca (Gd), while stage ii and $Ce_2Ni_7$ stage ii refer to the thickening stage of Mg-Ca (Gd) and $Ce_2Ni_7$ systems. (b-e**)** Energy landscape of the Mg-Gd, Mg-Ca, and $Ce_2Ni_7$ systems as a function of the normalized collective displacement $\xi_x$ and dissociative displacement $\xi_y$, with the unit of eV/Cell and the final relaxed structures' energy being the reference. The initial unrelaxed and final relaxed structures are denoted by red and black cross stars, respectively, and black curves represent the minimum energy path (MEP) of phase transformation.

TCP phases are formed by $M_R$-type hcp-orderings, implying that the undulating-$T$ net always constitutes the TCP/matrix interface (coherent terrace). Additionally, the $AB_2$ composition of the undulating-$T$ net indicates that the onset of dissociative displacements $\xi_y$ requires the presence of large $A$ atoms surrounded by small $B$ atoms. In the $Mg_2Ca$ case, the principal driving force for layer-separation can be attributed to the difference in atomic radii between Ca (1.8 Å) and Mg (1.5 Å)[58]. The $\xi_y$ prevents high strains in the basal planes and, when coupled with collective displacement $\xi_x$, leads to the formation of TCP structures. Hence, the formation of $MgZn_2$ phase requires Mg atoms to be surrounded by Zn atoms to create a local $AB_2$-type environment, thereby resulting in a larger size of $M_R$-type Zn ordering shown in Fig. A4. In addition, the type of hcp-matrix is crucial to the stability of $M_R$-type hcp-orderings. For example, in a separate study regarding high-throughput screening of TCP plates in hcp-metallic systems [59], we indicate that $M_R$-typ Ca orderings exhibit stability within hcp-Ti and hcp-Zr matrices without causing any structural transformations.

## 4.2. Critical nucleus in precipitation

The hcp→TCP structural transformations start from unstable hcp-orderings, however, not all of these orderings can evolve into a thermodynamically stable nucleus. According to the CNT, only when the size $n$ ($r$) of the nucleus is greater or equal to the size of the critical nucleus (i.e., $n(r) \geqslant n^*(r^*)$) can these nucleus continue to grow stably, otherwise that small-sized nucleus of $n(r)<n^*(r^*)$ will dissipate. Based on the structural evolution pathway of TCP plates proposed above, we here employed CNT to predict the critical sizes $n^*$ and nucleation barriers $\Delta G^*$ for the homogeneous nucleation of $Mg_2Ca$ and $MgZn_2$ phases, respectively.

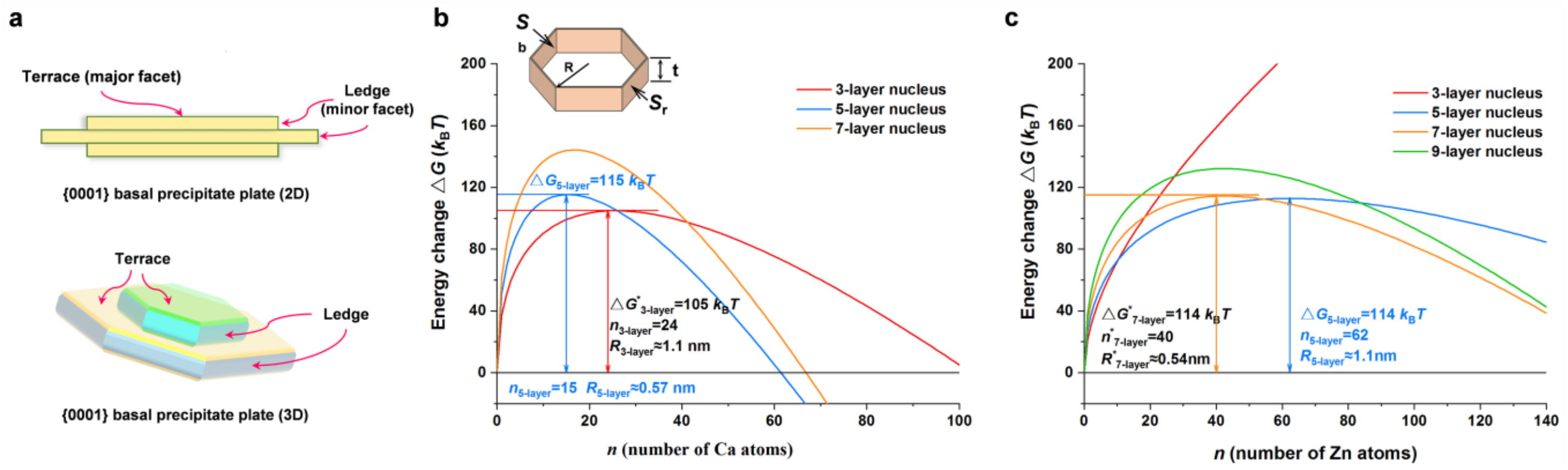

**Fig. 9.** The mesoscopic morphology of TCP precipitates and the CNT calculations about homogeneous nucleation for $Mg_2Ca$ and $MgZn_2$ precipitates. (a) The plate-shaped morphology of TCP precipitates in the hcp matrix. (b-c) Theoretical predictions of critical sizes $n^*$ and nucleation barriers $\Delta G^*$ in the homogeneous nucleation of $Mg_2Ca$ and $MgZn_2$ phases. The unit of energy change $\Delta G$ is $k_BT$, where $k_B$ is the Boltzmann constant and $T$=473.15 K.

The energy change $\Delta G$ associated with the nucleation of a precipitate with a volume $V$, surface area $S$ (including the areas of broad interface $S_b$ and rim interface $S_r$), and interfacial energies $\gamma$ (including $\gamma_b$ and $\gamma_r$) can be expressed as follows:

$$\Delta G = V\Delta G_{chem} + (S_b\gamma_b + S_r\gamma_r) + \Delta G_{elastic} \tag{6}$$

Where $\Delta G_{chem}$ represents the chemical free energy difference between the matrix and precipitate phase, while $\Delta G_{elastic}$ signifies the elastic energy within an infinite matrix containing a precipitate. According to previous experimental observations [22,44,46,57], TCP precipitates often display a hexagonal plate-like shape, as shown in Fig. 9a. Therefore, we assume that the nucleus is in the form of equilateral hexagonal plate (inset in Fig. 9b), with its radius of $R$ and the height of the precipitate rim being $t$. For the expression of $\Delta G_{Mg_2Ca}$ and $\Delta G_{MgZn_2}$, we have simplified Eq. (6) and their detailed calculations and discussion can refer to the Section 2.4.

Regarding the possible structures of the critical nucleus, the nucleus with different thicknesses is considered, respectively. Instead of using the typically employed $R$ as the independent variable[14,36] to plot $\Delta G$, we here plot the $\Delta G$ *vs*. $n$ (number of solute atoms in the nucleus) to facilitate comparing the critical sizes of nuclei with different thicknesses. In this way, Fig. 9b shows that the 3-layer $Mg_2Ca$ nucleus has the lowest critical nucleation energy barrier ($\Delta G^*_{3\text{-layer}}$=105 $k_BT$, with $n_{3\text{-layer}}$=24 Ca atoms, $R_{3\text{-layer}}$≈1.1 nm), but the 5-layer nucleus possesses the smallest size ($n_{5\text{-layer}}$=15 Ca atoms, $R_{5\text{-layer}}$≈0.57 nm, $\Delta G_{5\text{-layer}}$=115 $k_BT$). From an energy barrier perspective, the 3-layer nucleus is more likely to represent the critical nucleus over the 5-layer nucleus. However, the difference in nucleation energy barrier between the 3-layer nucleus and the 5-layer nucleus is relatively small, suggesting that both structures may appear in a real system.

In the case of $MgZn_2$ nucleation (Fig. 9c), similarly, both the 5-layer nucleus and 7-layer nucleus exhibit the same nucleation energy barrier ($\Delta G^*_{7\text{-layer}}$=$\Delta G^*_{5\text{-layer}}$=114 $k_BT$), but the critical size of the 7-layer nucleus ($n^*_{7\text{-layer}}$=40 Zn atoms, $R^*_{7\text{-layer}}$≈0.54 nm) is smaller than that of the 5-layer nucleus ($n_{5\text{-layer}}$=62 Zn atoms, $R_{5\text{-layer}}$≈1.1 nm). As a result, the 7-layer nucleus is predicted to be the energetically and kinetically preferred nucleus for $MgZn_2$ nucleation. And the structure of the critical nucleus is not fixed, but related to the type of TCP phase. As observed in the 3-layer structured $\gamma''$ phase[43,44] in Mg alloys, its critical nucleus should possess a 3-layer *M*-type structure. Admittedly, these predicted results need to be verified by more advanced experimental technologies in the future.

### 4.3. Conclusions

In conclusion, using first-principles calculations, we revealed the atomic-level nucleation and growth processes of TCP precipitates in hcp-Mg alloys, as well as the nonclassical nucleation behavior for this hcp-to-TCP structural transformation. Due to the similarity in lattice configuration between TCP and the parent hcp structures, the hcp→TCP transitions are considered conservative transformations, and the entire process is exclusively accomplished by the shuffle-based displacements. Moreover, we found that the unstable 3-layer $M_R$-type hcp-ordering acts as the basic structural transformation unit throughout the entire structural evolution, which results in the formation of TCP plates starting from a three-layer structure and thickens with two-layer as the thickness unit. During the thickening of TCP plates, the ledge nucleation depends on forming 3-layer $M_R$-type orderings on the coherent terrace of TCP plates, and the ledge does not exhibit the dislocation characteristics typically considered in the conventional perspective.

Furthermore, using CNT, the atomic structure of the critical nucleus is predicted to be associated with the type of TCP phases, with different thicknesses that may occur during precipitation. Overall, by locally sampling orderings structures, we demonstrate that ab initio calculations can accurately capture the basic structural transformation unit throughout the whole structural evolution process, and show that this basic structure governs the overall structural evolution. This provides great potential for utilizing ab initio simulations to explore the transformation of technically important coherent precipitates in other alloy systems, such as fcc-based Al alloys. Meanwhile, the insights gained are expected to shed light on the intricate processes of nucleation and growth at the atomic level and facilitate the design of novel materials containing TCP phases.

## Appendix A. Nucleation and growth processes of Laves phases

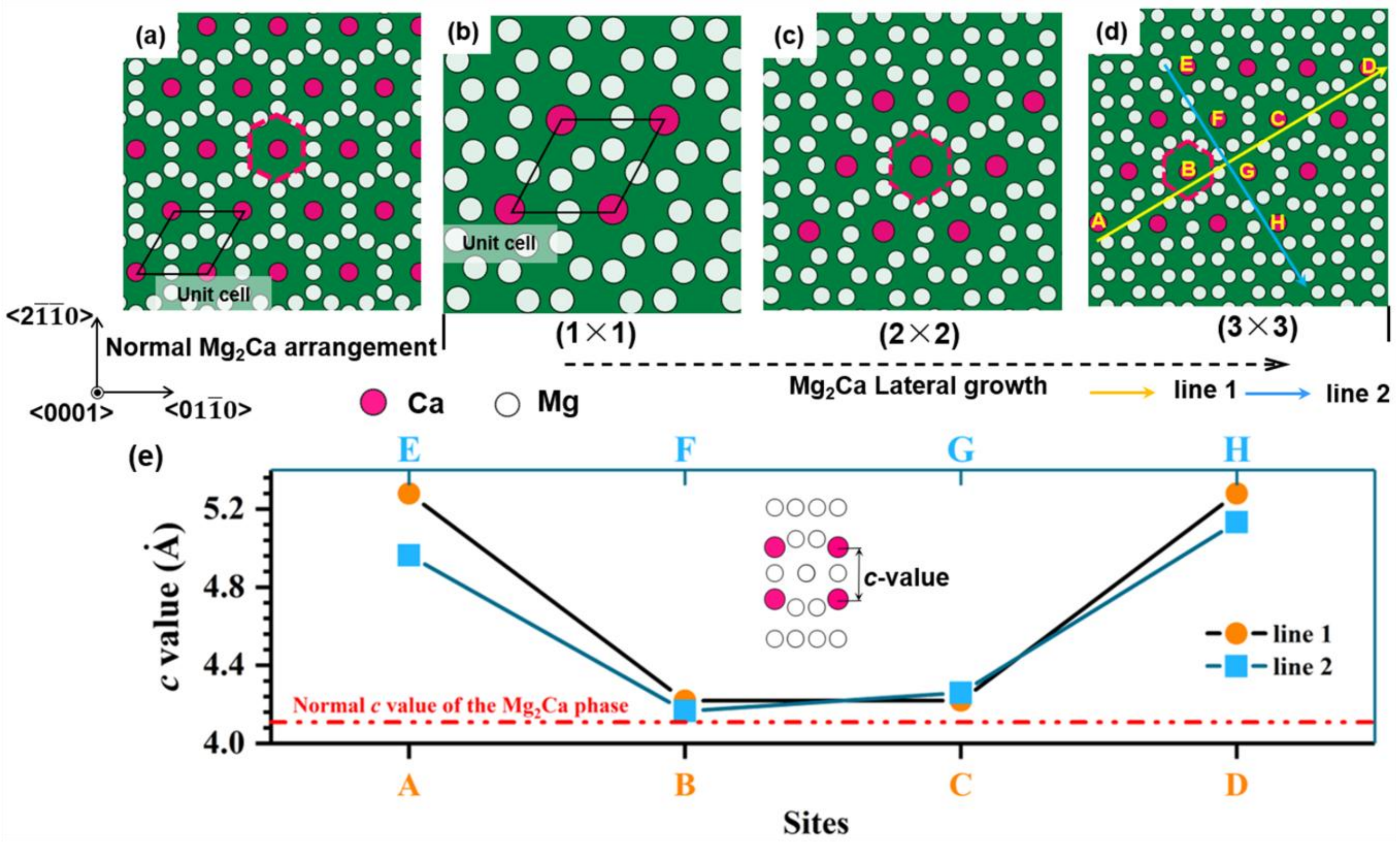

**Fig. A1.** (a) A projection of a normal *M*-type Ca ordering along with the <0001> direction. (b-d) The lateral growth of *M*-type Ca ordering within the Mg matrix. The unit-cell structure is marked as a black rhombus in panels (a) and (b). Ca and Mg atoms are plotted with pink and white circles, respectively. (e) The variation of the *c* value of *M*-type Ca ordering along the line 1 and line 2 directions that shown in panel (d). The horizontal dotted line highlights the normal *c* value of the $Mg_2Ca$ phase. A 450-atom model was used to perform this calculation.

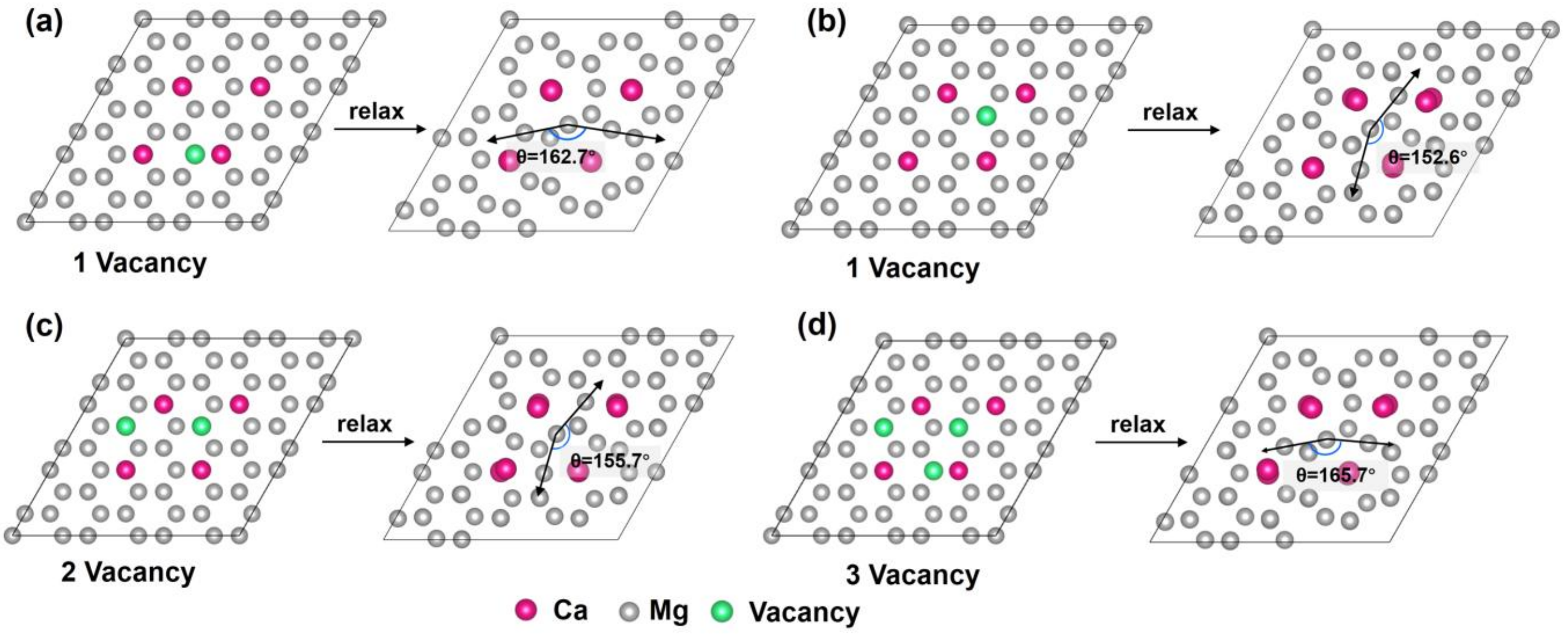

**Fig. A2.** The influence of different numbers of vacancies inside or nearby $M_R$-type Ca ordering on hcp-to-TCP structural transformations. The pink, grey and light-green balls represent the Ca, Mg atoms and vacancies, respectively.

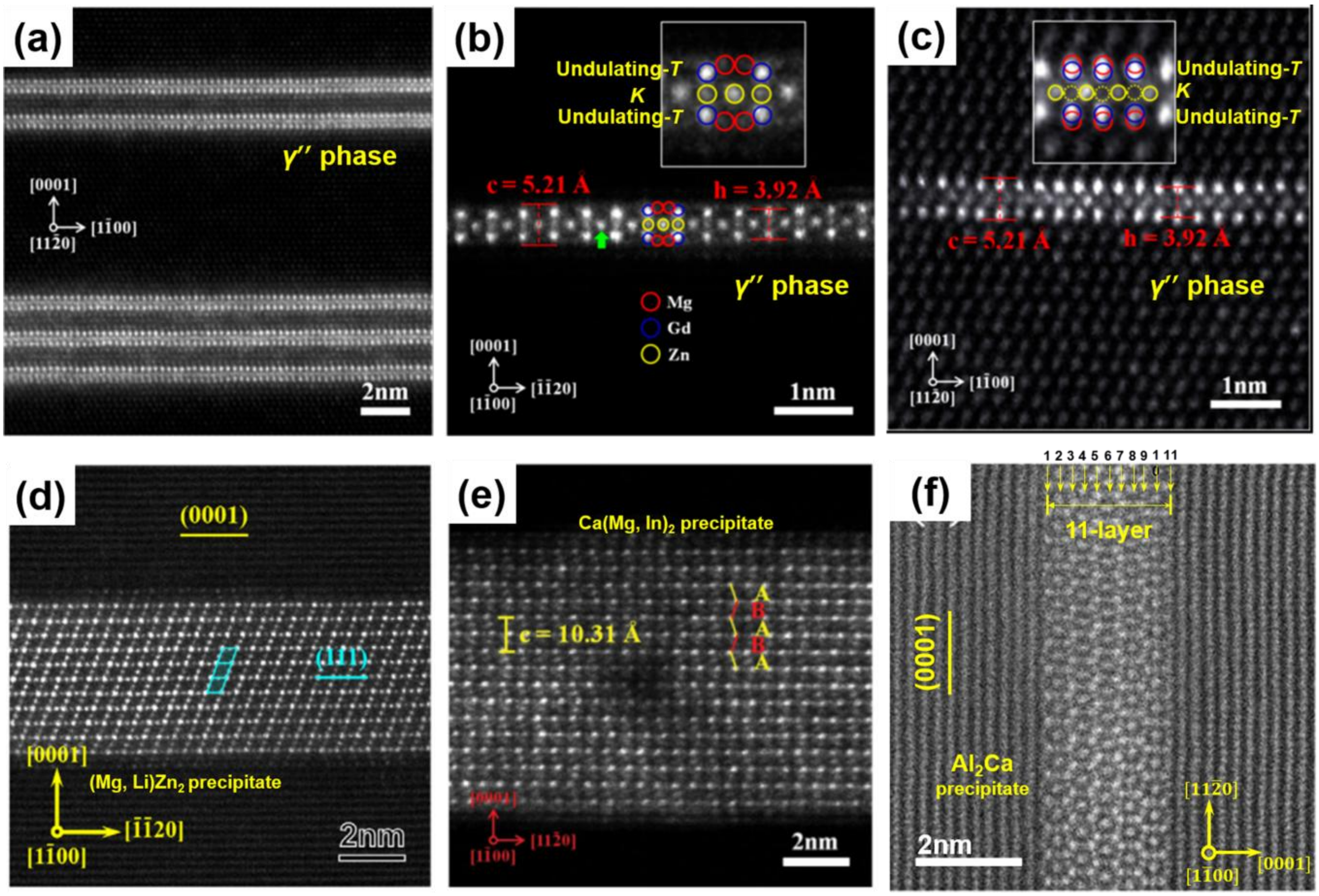

**Fig. A3.** Atomic-scale HAADF-STEM images of (a-c) γ′′ phase[44], (d) (Mg, Li)$Zn_2$[22], (e) Ca(Mg, In)$_2$[57] and (f) $Al_2Ca$[21] Laves phases. Panels (a-c) show γ′′ phase has a $CaCu_5$-*split* structure that is stacked with undulating-*T*, *K*, and undulating-*T* lattice nets. And (f) shows that the $Al_2Ca$ precipitate in the Mg matrix has 11-layer lattice nets, as indicated in the panel.

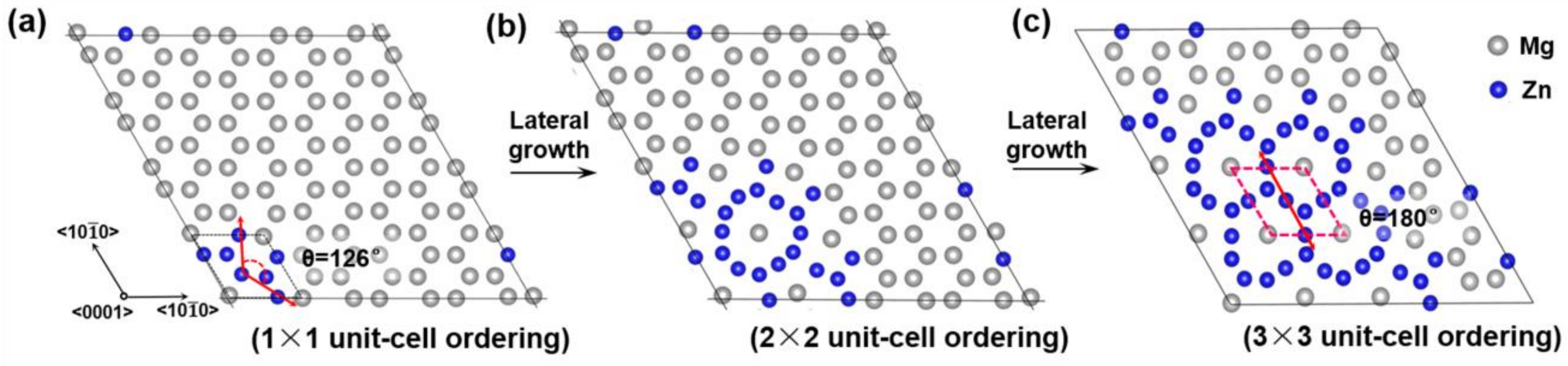

**Fig. A4.** The relaxed atomic structure for various sizes of $M_R$-type Zn ordering in the Mg matrix. We used a 450-atom model to perform these calculations. All structures are viewed in the direction of <0001>. The grey and dark balls correspond to the Mg and Zn atoms, respectively.

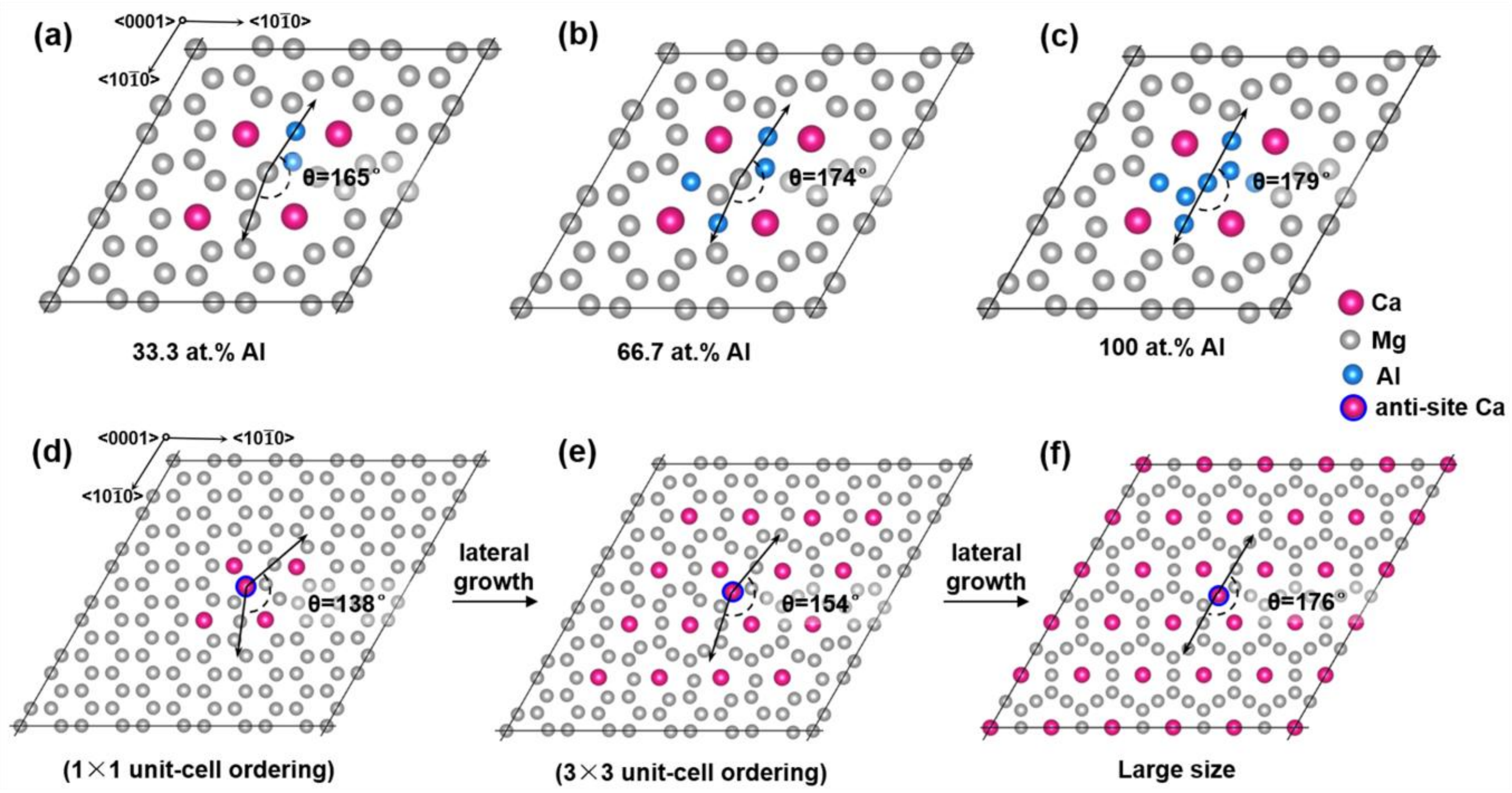

**Fig. A5.** Schematic diagram of non-classical nucleation of $Al_2Ca$ phase (a-c) and $Mg_2Ca$ phase (d-f) in Mg matrix. Within a unit-cell $M_R$-type ordering, nucleation of the $Al_2Ca$ phase utilizing (a) 33.3%, (b) 66.7%, and (c)100% atomic fraction of Al atoms. Despite one anti-site Ca atom residing in a unit-cell $M_R$-type ordering shown in (d), this ordering structure containing one anti-site Ca can still gradually transform into a normal *M*-type structure as the ordering size grows (e-f). All of the panels are oriented in the <0001> direction. The pink, grey, light-blue, and blue-edged pink balls represent the Ca, Mg, Al, and anti-site Ca atoms, respectively. The calculations in (a-c) and (d-f) were performed using the 162-atom and 450-atom models, respectively.

## Acknowledgments

This research is supported by the Fundamental Research Funds for the Central Universities of China (N2102011, N2007011, N160208001), 111 Project (B20029), and China Postdoctoral Science Foundation (2020M670774). We gratefully acknowledge Professor En Ma from Xi'an Jiaotong University for his invaluable comments and discussion on this work, and improvement of the manuscript.

## Author contributions

G. Qin and J. Bai conceived the original idea and designed the work. J. Bai conducted the simulations with help from X. Pang. All authors wrote the paper and discussed the results. G. Qin supervised the project.

**Declaration of Competing Interests**: The authors declare that they have no known competing financial interests or personal relationships that could have appeared to influence the work reported in this paper.

Supplementary Information for

# Structural pathway for nucleation and growth of topologically close-packed phase from parent hexagonal crystal

Junyuan Bai[1], Hongbo Xie[1,2] ,Xueyong Pang[1*], Min Jiang[1,3], Gaowu Qin[1,3*]

[1]*Key Laboratory for Anisotropy and Texture of Materials (Ministry of Education), School of Materials Science and Engineering, Northeastern University, Shenyang 110819, China*

[2]*State Key Laboratory of Rolling and Automation, Northeastern University, Shenyang 110819, China*

[3]*Research Center for Metal Wires, Northeastern University, Shenyang 110819 China*

**Content:**

## 1. Supplementary methods

**Binding energy**:

The binding energy $E_{bind}$ of a single solute atom X at different sites near the TCP precipitate is defined as the difference between a solute atom residing on the precipitate site and the bulk site. This energy can be expressed as:

$$E_{bind}=\left(E_{TCP}^{X}+E_{bulk}\right)-\left(E_{TCP}+E_{bulk}^{X}\right) \quad \text{(S-1)}$$

where the subscript indicates the structure of the host material, and the presence or absence of the superscript indicates whether or not a host atom is replaced with a solute atom X. For $E_{bulk}$ and $E_{bulk}^{X}$, we employed a 96-atom super-cell to determine the energy to eliminate periodic interactions of solute X. Based on this expression, it is clear that the negative energy means favorable binding thermodynamically, and a higher negative value indicates a stronger tendency of the site replacement by a solute atom.

**Convex hull construction**:

The convex hull (also designated as ab initio phase diagram at 0 K) method with energy input from first-principles calculations has been proven to be a powerful tool to explore the structures and stability of phases in various systems[1–3]. In this work, a convex hull was constructed with $Mg_2Ca$ and $Al_2Ca$ phases as two ends to qualitatively evaluate the impact of composition on the structure of intermediate $(Mg_{1-x}, Al_x)_2Ca$ phases. The MAPS code in the alloy theoretic automated toolkit (ATAT)[4] was adopted to enumerate hypothetical configurations with varying fractions of Al in terms of the C14, C15, and C36 structures. All structures with up to 24 atoms/unit-cell were searched in the current work.

The formation enthalpy $H_f$ of a configuration $A_xB_y$, for instance, was calculated relative to the zero kelvin total energies of pure elements A and B as follows:

$$H_f=\frac{E\left(A_xB_y\right)-N_xE_A-N_yE_B}{N_x+N_y} \quad \text{(S-2)}$$

Where $E(A_xB_y)$ denotes the total energies of a configuration, and $E_A$ and $E_B$ are the total energies per atom for pure elements $A$ and $B$, respectively. $N_x$ and $N_y$ represent the number of elements $A$ and $B$ in the configuration. In this way, the formation enthalpy of a configuration with ternary or more constituents can be obtained.

## 2. Supplementary texts-I: Discussions about atomic models utilized in the DFT calculations

In their earlier study, as shown in Fig. S1, Natarajan et al.[5] adopt a model without containing the matrix portion (i.e., a matrix-free model) to study the hcp→TCP transformations. As their results show, upon relaxation, the TCP structures, like the C15-type structure, can be directly transformed from their constructed models. Although their models do generate TCP structures, their modeling approach is extremely unreasonable. Typically, the nucleation of a nanoprecipitate starts from certain nano-sized clusters[6,7], implying that these clusters (nucleus) exist in an aperiodic environment surrounded by the matrix. Because the matrix-free model by default disregards this aperiodic scenario, it fails to replicate the actual precipitation behavior of TCP phases.

On the other hand, employing a matrix-free model will ignore the impact of different hcp-matrices on the hcp-to-TCP transformation and lead to erroneous conclusions. For instance, we examine the possibility of the $MgZn_2$ Laves phase precipitating in situ from the hcp-Mg and Zn matrices, respectively. The matrix-free model can merely indicate that the combination of Mg and Zn can form Laves structures in this particular scenario, but cannot further predict that the $MgZn_2$ phase will only precipitate in the Mg matrix and not in the Zn matrix, as depicted in Fig. S2 and Fig. A4. Moreover, this matrix-free model only involves the nucleation stage, in which the hcp-ordering structure is infinite by default, and ignores the growth process, resulting in atomic details about nucleation and growth processes that remain unknown yet.

Therefore, in the current work, we use models that rigorously consider the matrix portion to resolve the formation pathways of TCP phases in the hcp-matrix. The results of supercell size converge tests (Fig. S3) show that the model utilized in the main text is sufficiently converged, with no notable change in lattice parameters as supercell size increases. Moreover, it is noteworthy that the other six new transition pathways recently reported by Kolli et al.[8], such as the hcp→C16 transformation, also employ matrix-free models to perform DFT calculations. These recently reported pathways require more examinations.

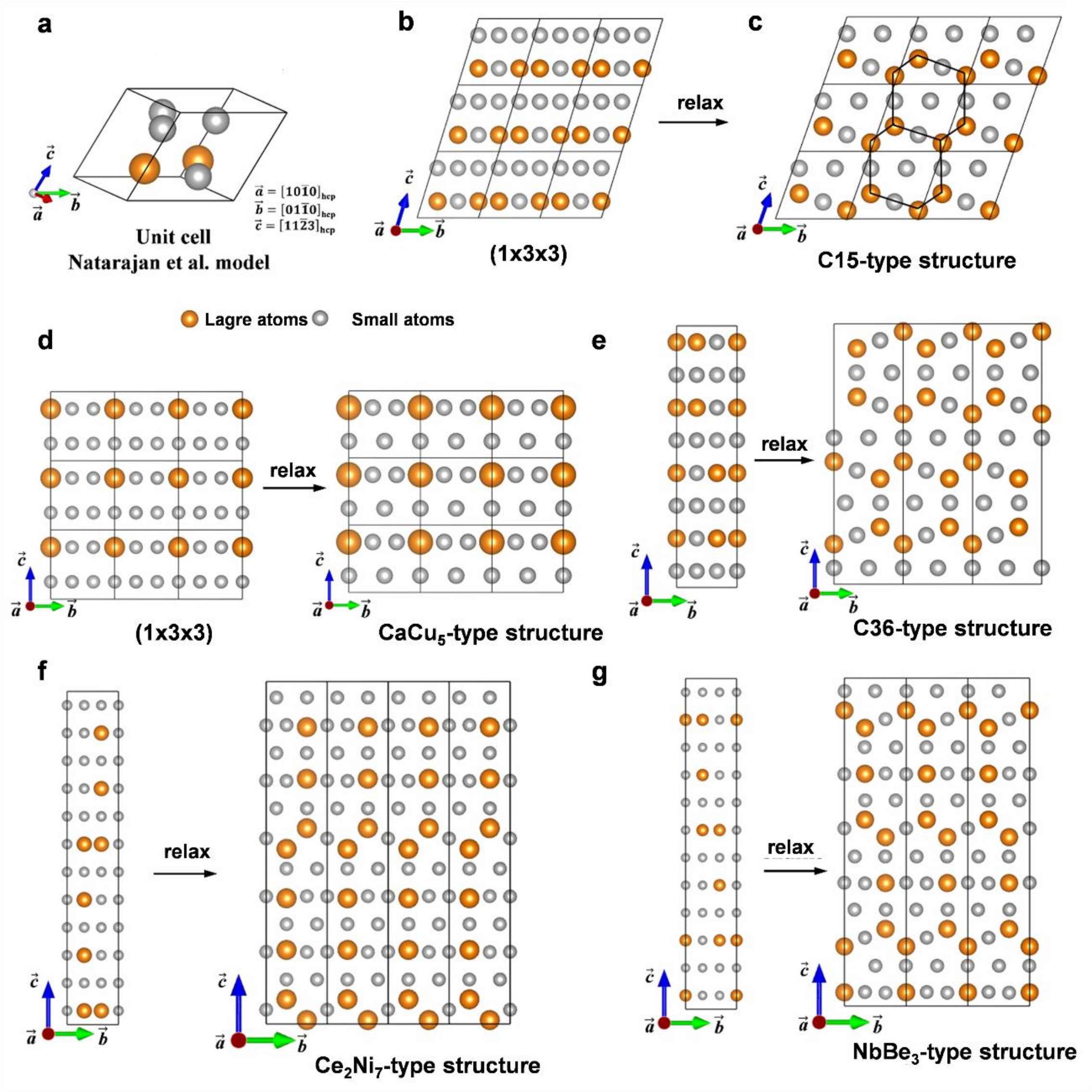

**Fig. S1.** A schematic depiction of the atomic models employed by Natarajan et al.[5]. (a-c) The formation process of a C15-type structure by a unit-cell hcp-ordering structure after DFT-relaxation. The lattice vectors $\vec{a}$, $\vec{b}$ and $\vec{c}$ parallel to the $[10\bar{1}0]$, $[01\bar{1}0]$ and $[11\bar{2}3]$, respectively, are used to construct the unit-cell model. (d-g) The formation processes of $CaCu_5$-, C36, $Ce_2Ni_7$- and $NbBe_3$-types structures by the unit-cell hcp-ordering structures after DFT-relaxation. These models are all established with their vectors $\vec{a}$, $\vec{b}$ and $\vec{c}$ parallel to the $[10\bar{1}0]$, $[01\bar{1}0]$ and $[0001]$, respectively. The black-framed parallelograms and rectangles denote the unit-cell model. Large and small atoms are colored with yellow and grey balls, respectively.

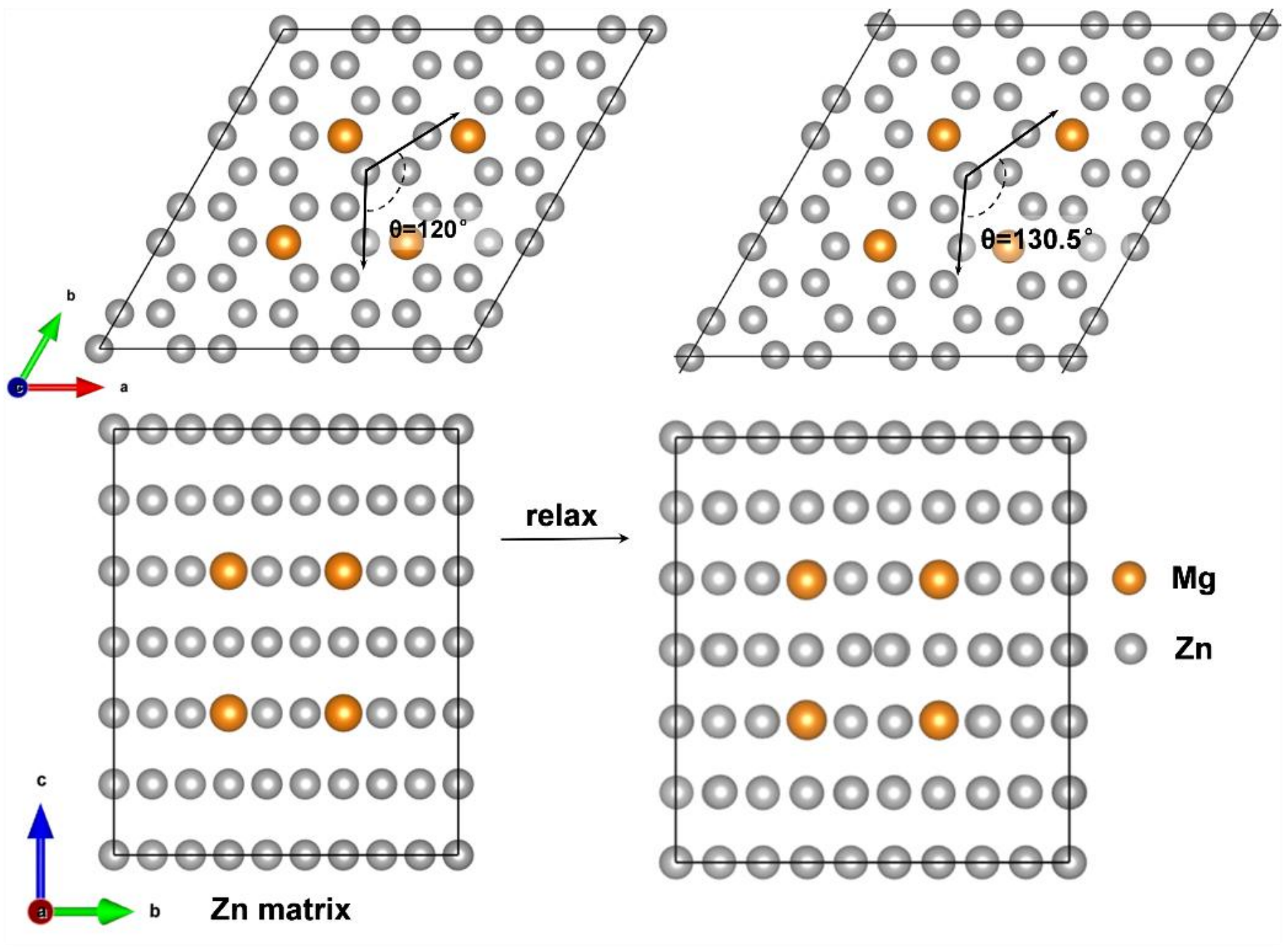

**Fig. S2.** The atomic structure of a unit-cell $M_R$-type Mg ordering in the Zn matrix after a structural relaxation. The θ value changes from 120° to 130.5°, while there is no evident split phenomenon like in the Mg matrix, indicating that the $MgZn_2$ phase is hard to form in the Zn matrix. The left side and right side of the panel exhibit the initial un-shuffled and shuffled structures, respectively. A 162-atom model is used in this calculation. The Mg and Zn atoms are represented by the yellow and grey balls, respectively.

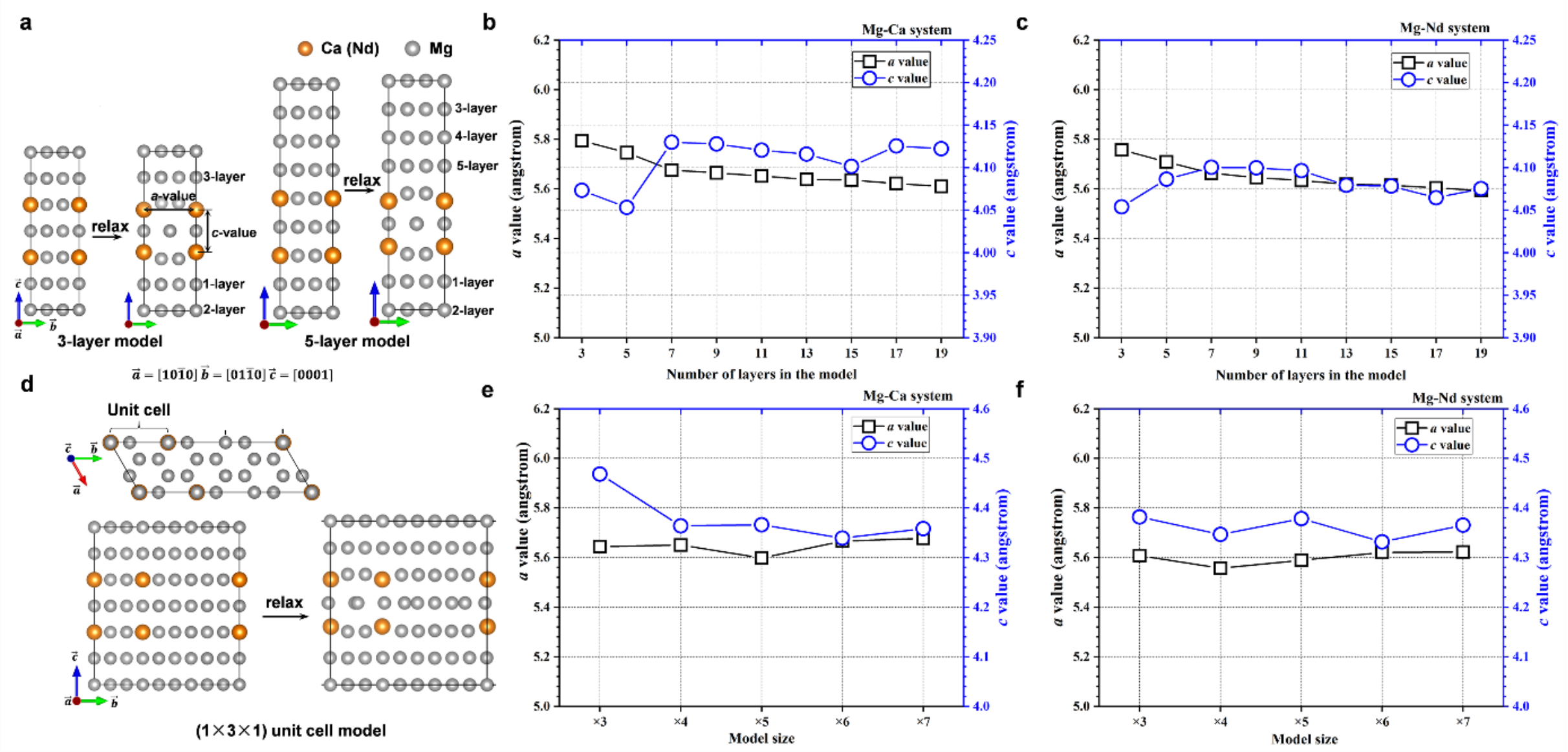

**Fig**. **S3**. Convergence tests for super-cell size. (a) Schematics of various sized atomic models were used to perform the convergence test along the *c*-axis. (b-c) The calculated *a* value and *c* value of relaxed structure as functions of the number of layers for Mg-Ca and Mg-Nd systems. (d) Schematics of different sized models used to perform the convergence test along the *b*-axis direction. (e-f) The calculated *a* value and *c* value of relaxed structure as functions of the sizes of the model for Mg-Ca and Mg-Nd systems. The Ca (Nd) and Mg atoms are represented by yellow and grey balls, respectively.

## 3. Supplementary texts-II: Crystallographic structures of Laves-like phases

In the present study, the $CaCu_5$-type, $Ce_2Ni_7$-type, $Y_2Ni_7$-type, and $NbBe_3$-type phases are categorized as Laves-like phases because they are made up of alternating triangular and kagome lattice nets. Different from Laves phases, these Laves-like phases generally contain the planar-*T* lattice nets within their structures. For instance, the $CaCu_5$-type phase consists of alternating planar-*T* and *K* lattice nets (Fig. S5(a-c)), while the $Ce_2Ni_7$-type phase contains the puckered-*T* and planar-*T* lattice nets (Fig. S4). Notably, the $Fe_7W_6$ $\mu$ and $Zr_4Al_3$-type phases can also be viewed as the Laves-like phases. However, these two phases still contain the ω-*T* lattice net ($\rho_{\omega\text{-}T} = 4$), which results in $T \rightarrow \omega\text{-}T$ transition necessitating additional interstitial atoms. Since the $\mu$ and $Zr_4Al_3$-type precipitates have not been observed in hcp metallic systems, this work does not take into account such non-conservative transformations.

Similar to the new definition for the crystal structure of Laves phases shown in Fig. 2c, the $Ce_2Ni_7$-type, $Y_2Ni_7$-type, and $NbBe_3$-type structures can be represented with stretched-building blocks (Fig. S4(d-f)), while the $CaCu_5$-type structure is exclusively constructed by the *M*-type subunit (Fig. S5(d)). These stretched-building blocks are likewise composed of five subunits, as shown in Fig. S5(e).

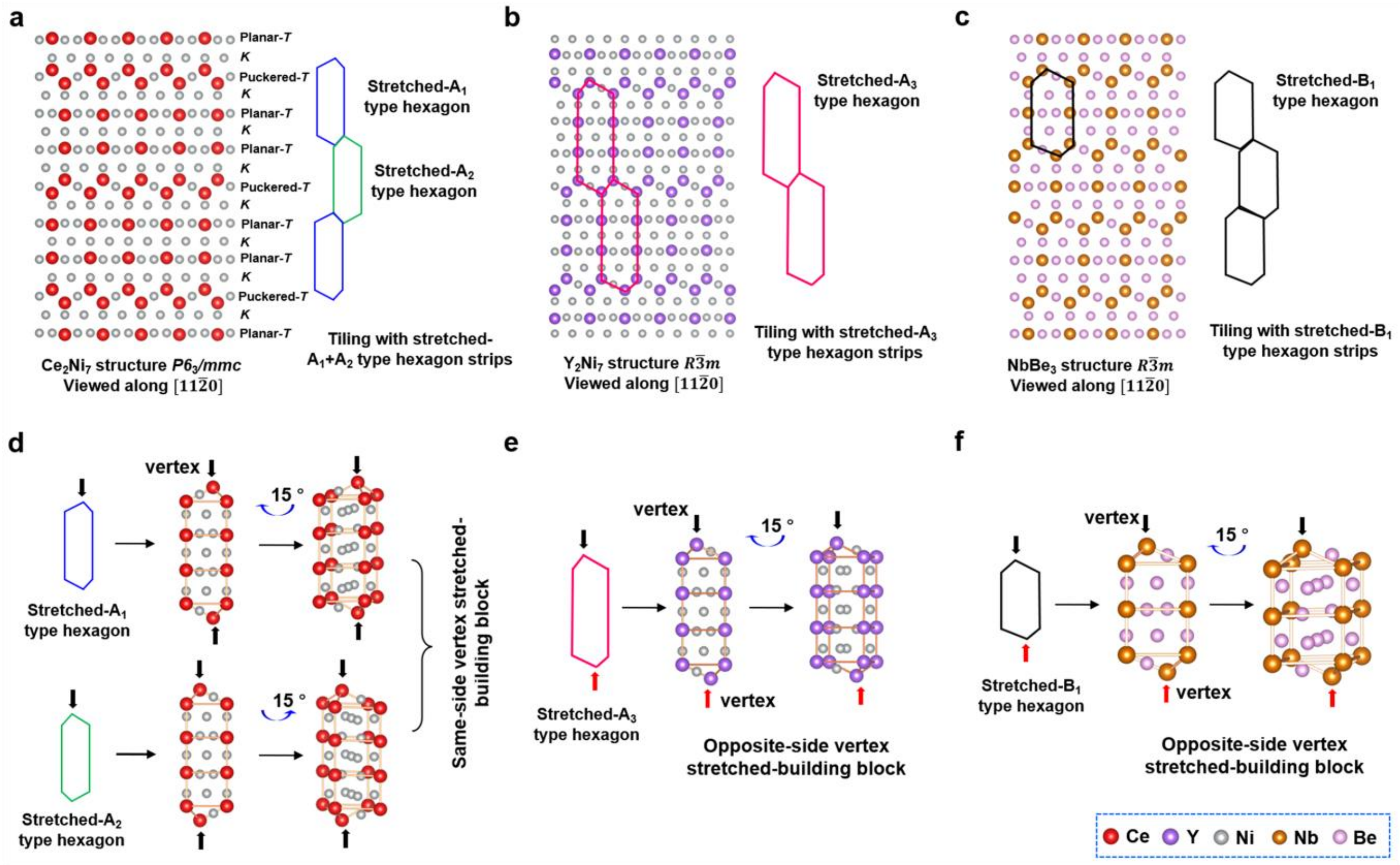

**Fig. S4.** The schematic diagram for assembling $Ce_2Ni_7$, $Y_2Ni_7$, and $NbBe_3$ phases with stretched-building blocks. (a-c) The [11$\bar{2}$0] projection of $Ce_2Ni_7$, $Y_2Ni_7$, and $NbBe_3$ phases can be viewed as tiling with four different types of stretched-hexagons, which are represented by blue-edged stretched-$A_1$ type, green-edged stretched-$A_2$ type, pink-edged stretched-$A_3$ type, and black-edged stretched-$B_1$ type hexagons, respectively. The red, purple, light-grey, yellow, and pink balls represent Ce, Y, Ni, Nb, and Be atoms, respectively. (d-f) The 3D crystallographic structure of four types of stretched-hexagons, where stretched-$A_1$ and stretched-$A_2$ type hexagons belong to the same-side vertex (indicated with black arrows) stretched-building block and the other two belong to the opposite-side vertex stretched-building block (labeled as black and red arrows).

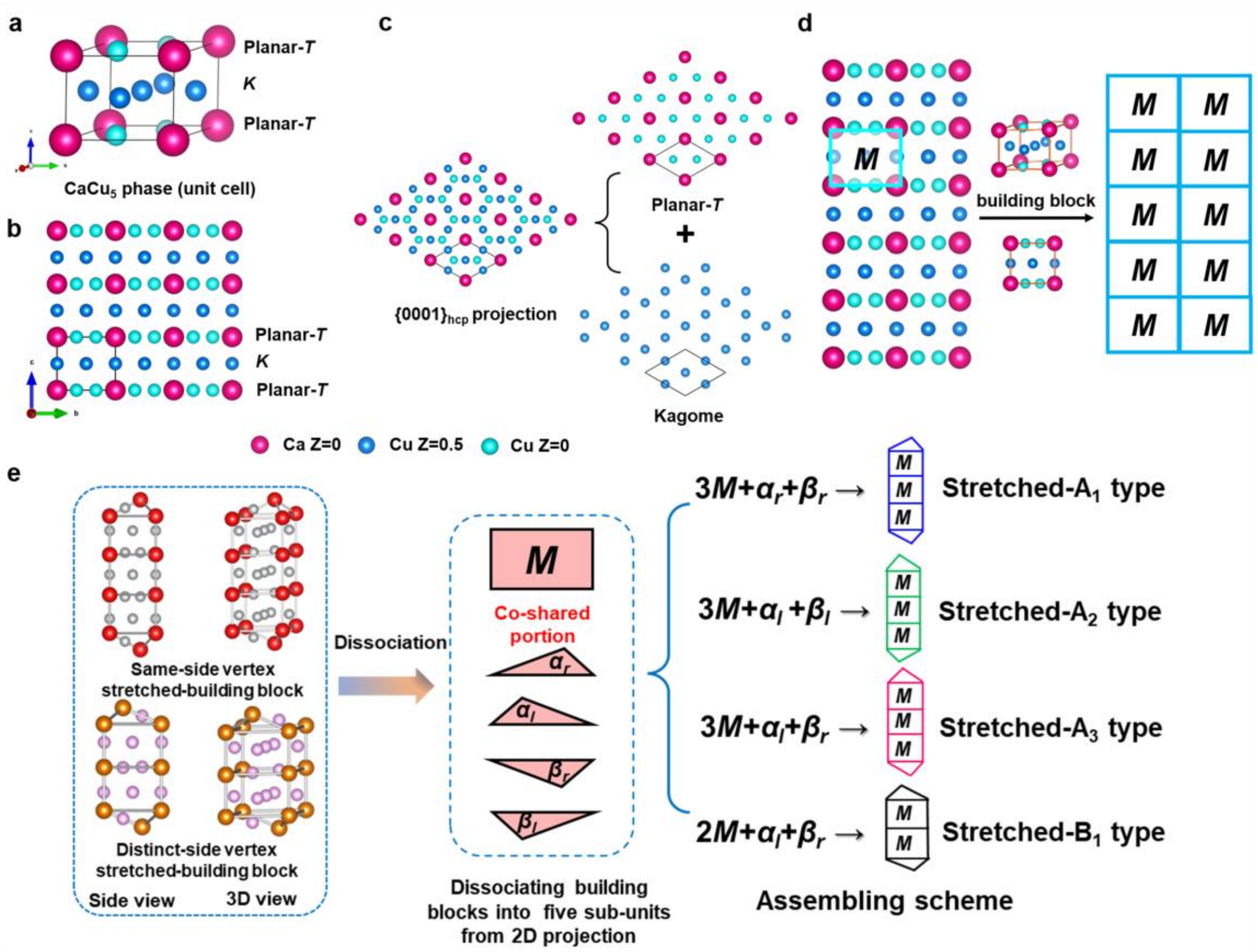

**Fig. S5**. (a-c) Schematic crystallographic structure for the $CaCu_5$ phase. The type of lattice net in this $CaCu_5$ phase is indicated by the notations on the right side of panel. And the dark frames in (b-c) indicate the $CaCu_5$ phase's unit cell structure. (d) Schematic illustration of assembling $CaCu_5$ phase using *M*-type structures. (e) Dissociative five subunits of building blocks in the $Ce_2Ni_7$, $Y_2Ni_7$, and $NbBe_3$ phases. The right side of the panel shows their combinational strategy to form four types of stretched-building blocks.

## 4. Supplementary texts-III: Structural evolution pathways for Laves-like phases within hcp-Mg matrix

An alternative route to address the restriction of the undulating-*T* net on *c*-axis periodicity involves dealing with the Laves-like structures. Typically, these Laves-like phases are used as the bulk form in the fields of hydrogen storage[9–11] and magnetic materials[12], whereas their existence in alloys has not been explicitly observed. Thus, in this work, we merely predicted the structural evolution pathways of these Laves-like phases within the hcp-Mg matrix and provided preliminary insights into why they are infrequent in alloys. Given their similar lattice structures, we selected the $Ce_2Ni_7$ phase (a thermodynamically stable phase in the Mg-Ce-Ni phase diagram[13]) as an example to illustrate the structural evolution pathway of Laves-like phases in the hcp-Mg matrix.

Apart from establishing $M_R$-type orderings on the $S_R$ or $S_L$ atomic sites, as shown in Fig. 3(b), these orderings can yet be constructed on their original substitutional atomic sites to thicken this $Ce_2Ni_7$ phase. After structural transformation, the undulating-*T* net becomes a planar-*T* lattice net (i.e., undulating-*T*→planar-*T*), indicating that previously outward small atoms are compressed back to the original plane. Besides, the undulating-*T*→puckered-*T* transition is also involved in the late thickening process of the $Ce_2Ni_7$ phase. Hence, these Laves-like phases can be thickened in this manner without changing the stoichiometry of the undulating-*T* nets. However, as illustrated in Fig. S7, the onset of the undulating-*T*→planar-*T* lattice transition is associated with the chemical composition of the undulating-*T* lattice net. The thickening of the $Mg_2Ca$ Laves phase is achieved solely by modifying the stoichiometry of its undulating-*T* lattice net, rather than forming Ca orderings on their original substitutional atomic sites. However, Laves-like phases can simultaneously employ two thickening routes for structure thickening, as Fig. S6(b) shows. Moreover, the undulating-*T*→planar-*T* lattice transition is fully completed when those Ni atoms completely occupy the $Ce_2Ni_7$ structure's small-atomic sites. Given the non-classical nucleation process involved in the formation of TCP phases, the undulating-*T* nets are probably hard to fully transform into a normal planar-*T* lattice net due to kinetic considerations in practical precipitate processes. Therefore, given the solubility of specific solutes, aging temperature, and other objective conditions, it may prove challenging to observe the typical structures of these Laves-like precipitates in alloys.

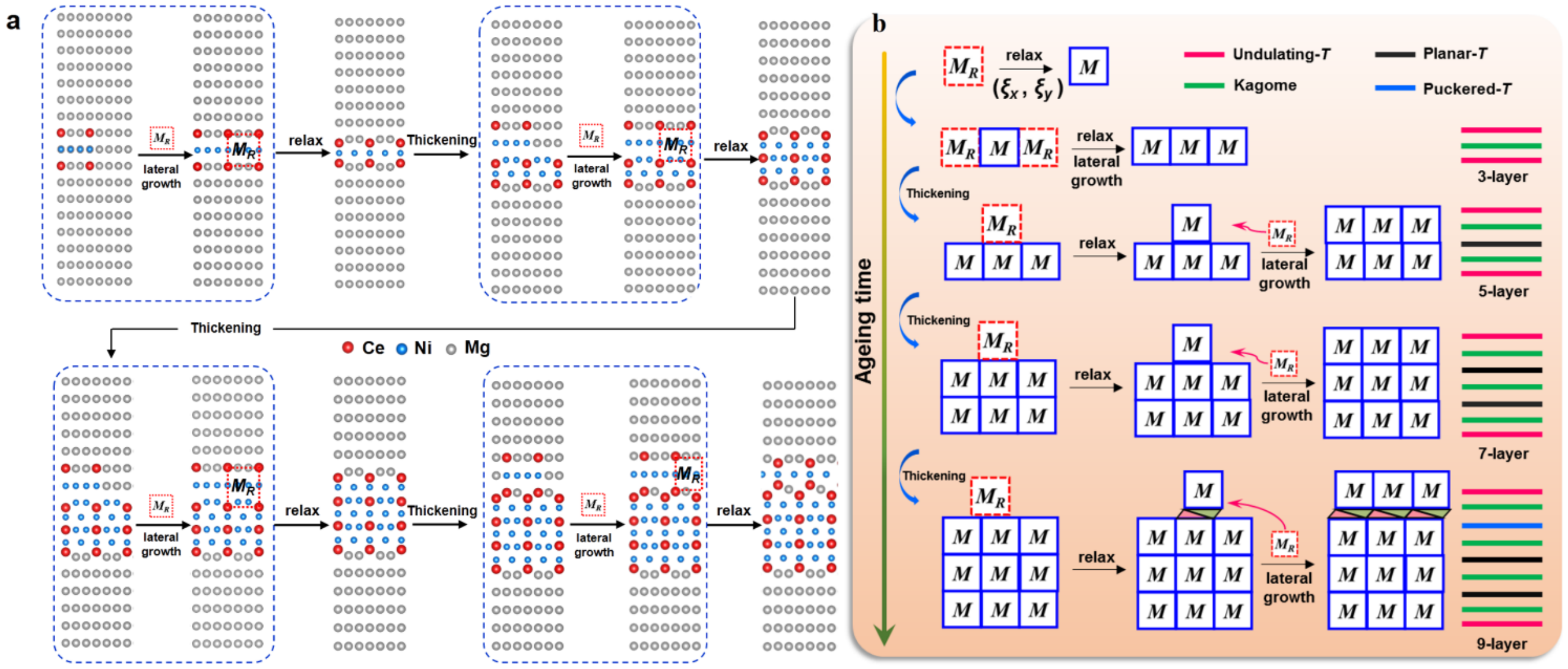

**Fig. S6**. (a) The initial nucleation and thickening processes of the $Ce_2Ni_7$ phase within the Mg matrix. Through building $M_R$-type orderings (red-dotted rectangle) on the newly formed TCP/matrix interface, the $Ce_2Ni_7$ phase formed in the Mg matrix. The Ce, Ni and Mg atoms are represented by the red, blue and grey balls, respectively. (b) The suggested formation pathway for the $Ce_2Ni_7$ phase.

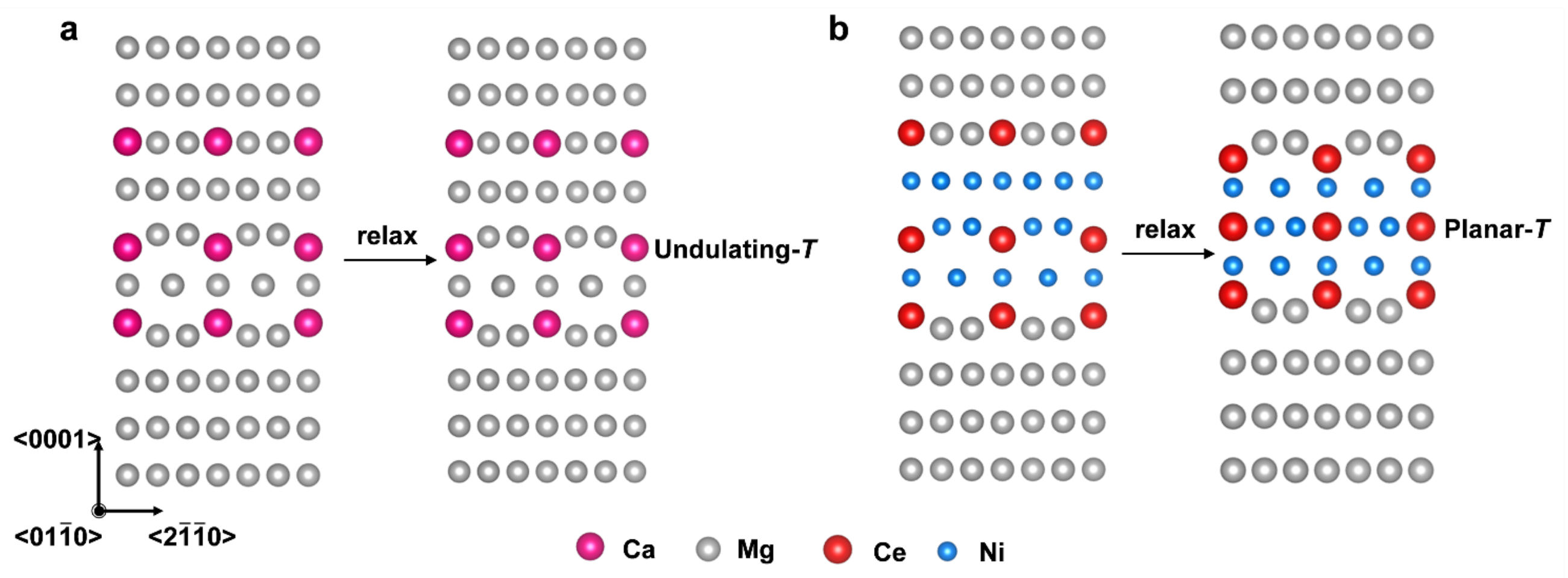

**Fig. S7.** (a) The result of thickening $Mg_2Ca$ phase using the same route as that of Laves-like phases. (b) The thickening process of the $Ce_2Ni_7$ phase, with the original undulating-*T* net transforms into the planar-*T* lattice net. The Ca, Mg, Ce and Ni atoms are marked with dark-pink, grey, red and blue balls, respectively.

## 5. Supplementary texts-IV: Plate-like morphology of TCP precipitates

In real systems, the evolution of microstructure is driven by energy minimization, i.e., the growth of TCP plates facilitates the decline of system energy after the nucleation stage. Although the broad interface of the TCP plates maintains a high degree of coherence with the Mg matrix, the larger lattice distortion induced by the formation of TCP structures around the edge of TCP plates is still inevitable, especially the strain caused by the formation of *K* net via collective displacement $\xi_x$, as shown in Fig. S8. Such lattice distortion exhibits an autocatalytic-growth effect for the lengthening process of TCP plates, as illustrated in Fig. S9. By calculating the binding energies (Eq. (S-1)) of solute atoms on different atomic sites (marked with capital letters) near the TCP plate interface, the tendency of solute atoms to occupy specific sites can be qualitatively evaluated. A negative binding energy indicates that a solute atom occupying this site can bring a system lower energy than in the bulk environment, leading to solutes energetically preferred to migrate to this position.

In Fig. S9(a-b), for lateral growth of the 3-layer and 5-layer TCP structures, the large Ca atoms exhibit the most negative values at the C-site, while the small Al and Zn atoms show the greatest negative values at the E-site, as indicated by the enclosed yellow areas. Meanwhile, the large Ca atoms on the E-site display the highest positive binding value. This indicates the preference for Ca and Al (Zn) atoms to migrate towards the C-site and E-site, respectively, from the matrix to facilitate the lengthening process of TCP plate. However, another dissociative displacement $\xi_y$ does not produce an obvious attractive effect on the solutes migrating toward the coherent terrace, due to positive binding values or oscillating around 0 eV (Fig. S9(c)). This leads to ledge nucleation appears to be a stochastic process, lacking a pronounced driving force comparable to that of lateral growth. Consequently, due to rim and broad interfaces being very different in attracting solute atoms, most TCP precipitates prefer to evolve plate-like shapes within the hcp matrix.

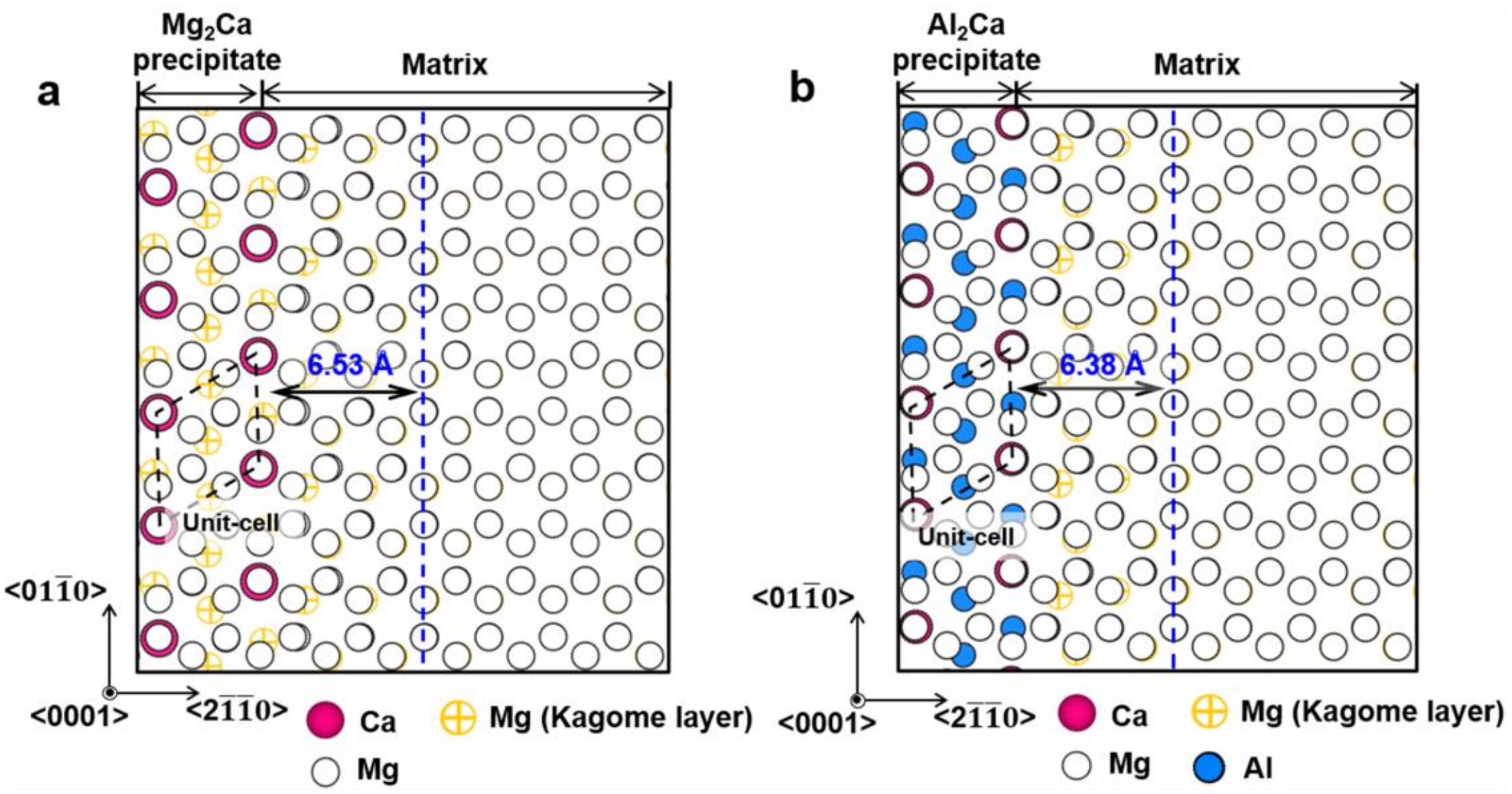

**Fig. S8.** The lateral side facets' projection of (a) $Mg_2Ca$ precipitate and (b) $Al_2Ca$ precipitate along the <0001> direction. The unit-cell structure of the precipitate is circled with a black-dotted rhombus on the left side of the panel. And the range of strain-affected zone produced by forming $Mg_2Ca$ ($Al_2Ca$) precipitates is also indicated in the panel.

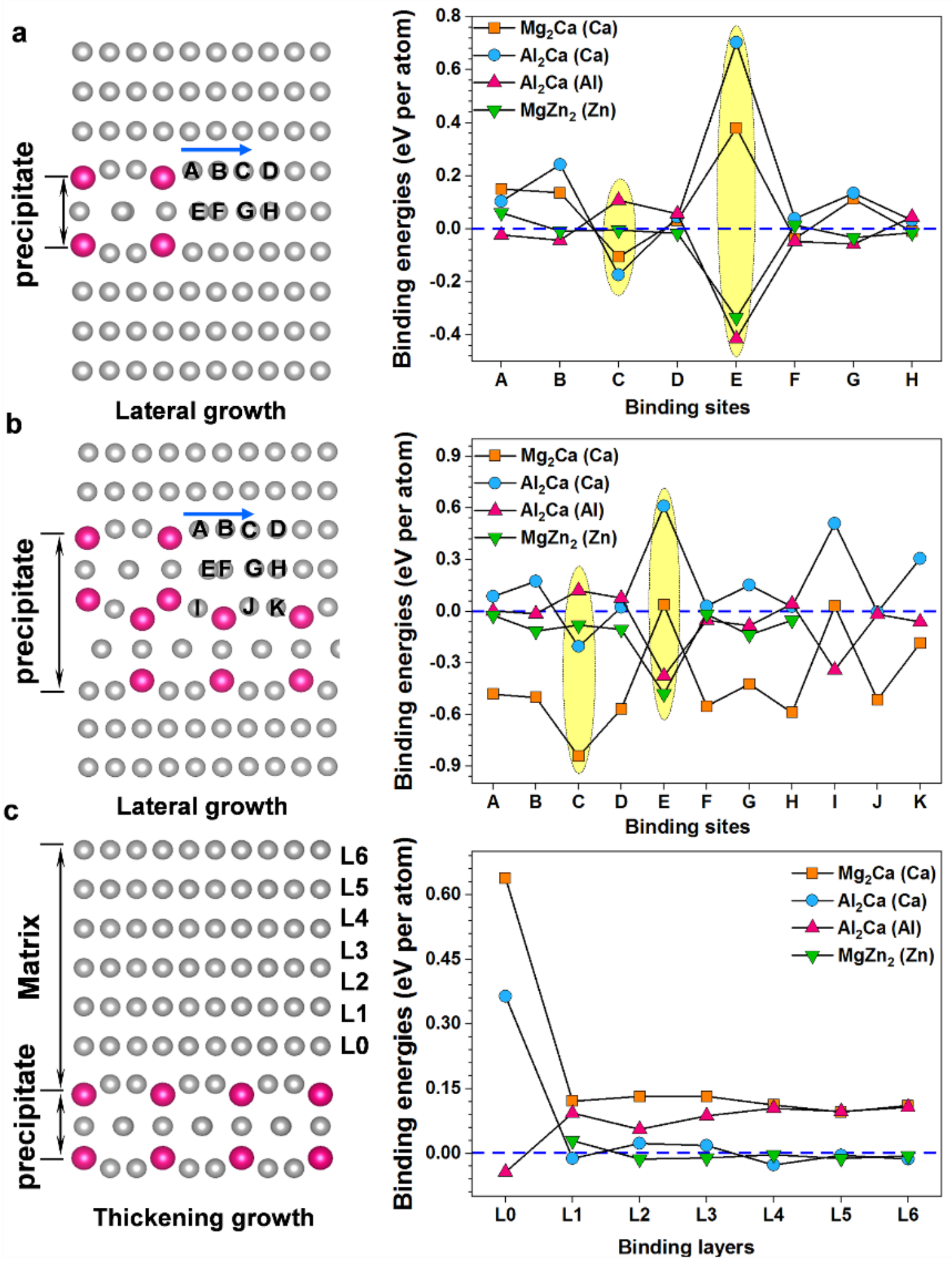

**Fig. S9.** Solute (given in parentheses) binding energies on distinct atomic sites near the lateral side facet of (a) one-unit-cell height precipitate and (b) thickened precipitate. (c) Solute binding energies near the coherent interface on various atomic basal layers. The different atomic sites near the precipitate/matrix interface are shown on the left side of each panel. And the large and small atoms are represented by the dark-pink and light-grey balls, respectively.

## 6. Supplementary texts-V: Polytypes and off-stoichiometry phenomena in TCP precipitates

During the precipitation process, the actual TCP phases in alloys often display various intricate phenomena, including off-stoichiometry and polytypes in structure. These features are especially noticeable in Mg-free TCP phases, such as the $Al_2Ca$ phase [7,14,15]. Using the $Al_2Ca$ phase as an example, we found that these features basically arise from nonclassical nucleation behaviors and are sensitively influenced by the interplay between thermodynamics and kinetics. As mentioned above, the $Al_2Ca$ phase forms through a nonclassical nucleation process, wherein the onset of structural transformation depends exclusively on the distribution of Ca atoms in the nucleus rather than that of Al atoms. Furthermore, several experimental and theoretical studies [16–18] demonstrate that Ca is a rapidly diffusing element in the Mg matrix, with its diffusion rate being approximately 2~3 orders of magnitude higher than that of Al at ~500 K. Due to such significant discrepancies in bulk diffusion, it is unlikely for $Al_2Ca$ precipitates to achieve a normal stoichiometry, instead, off-stoichiometry phenomena are readily observed in precipitates.

Besides causing the deviation in stoichiometry, the kinetic factors can also lead to polytypes in precipitates. In Fig. S10(a), the Mg-Al-Ca ternary phase diagram (0 K) shows that the $Mg_2Ca$ and $Al_2Ca$ Laves phases are thermodynamically stable (i.e., equilibrium) phases, and their stable structures at low temperatures are C14 and C15 structures (Fig. S10(b)), respectively. However, achieving an ideal stoichiomerty for the $Al_2Ca$ phase might be hard due to the impact of kinetic factors, especially at at the early stages of precipitation. Fig. S10(b) shows that the formation enthalpies of the non-stoichiometric $(Mg_{1-x}Al_x)_2Ca$ compounds ($0<x<1$) all resides above the global convex hull, indicating metastable nature of these compounds. And the structure with the minimum convex hull distance on each composition is recognized as a possible metastable structure [19,20]. Therefore, C14 structures prefer to form at lower Al content ($x \leq 9/16$), whereas C36 structures become more probable within the composition range of $9/16<x\leq15/16$. In $(Mg_{1-x}Al_x)_2Ca$ compounds, these two structures have both been experimentally observed[14], lending credence to our prediction.

Simultaneously, one should note that our configuration enumeration is just based on C14, C15, and C36 structures, which leads to the limited size of the model used (Supplementary methods). Consequently, more intricate metastable states, such as a precipitate incorporating both C14 and C15 structures[21], have not been taken into account in the first-principles calculations due to the substantial model size. Moreover, Fig. S10(b) further shows the tiny energy disparity among diverse structures, indicating that these structures are thermodynamically competitive [22]. Given the wide range of synthesis variables, including alloy composition and aging temperature, the formation of complex metastable TCP precipitates is facilitated during the aging treatment. Overall, the formation of complex microstructure arises from the nonclassical nucleation mechanism of TCP precipitates, whether they are in equilibrium or metastable states, and is further influenced by the subtle interplay between thermodynamics and kinetics.

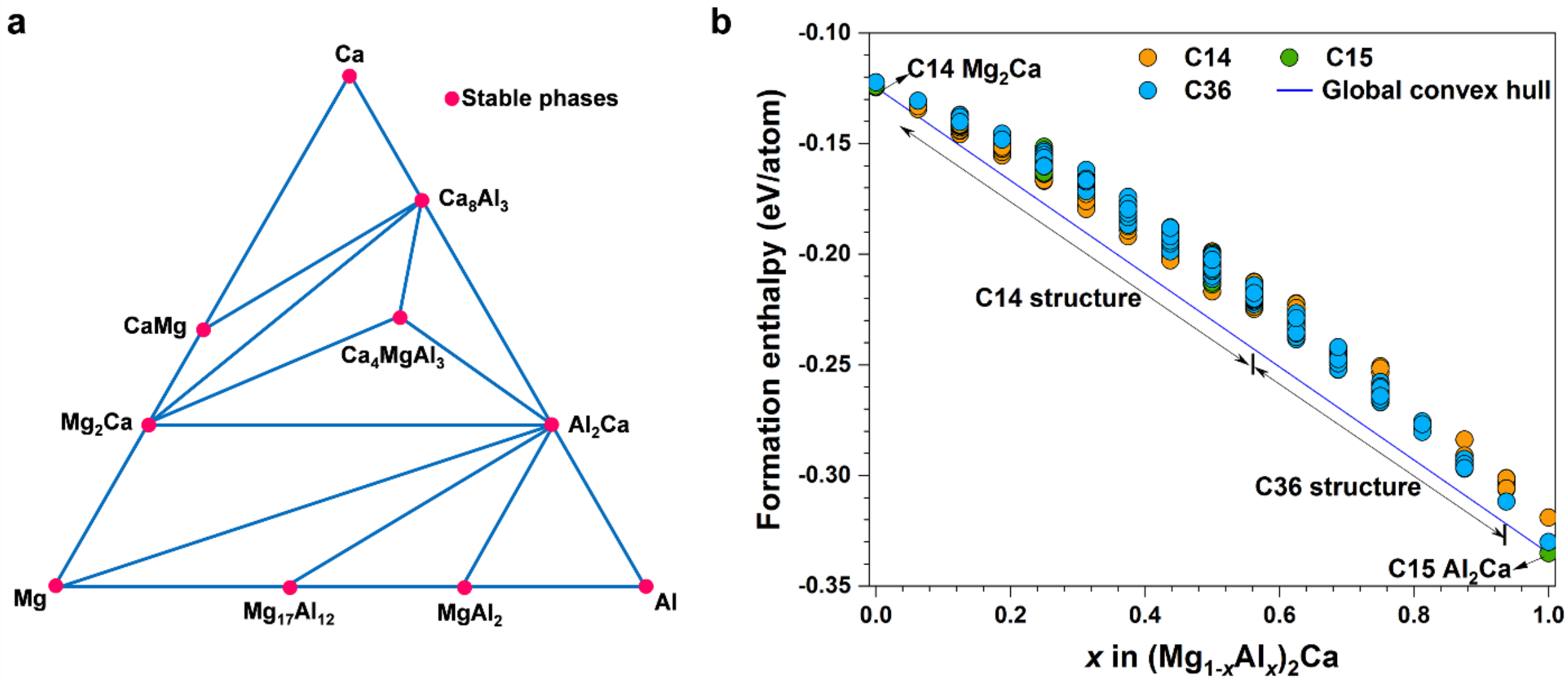

**Fig. S10.** (a) A phase diagram for the Mg-Al-Ca ternary system that was constructed based on the Materials Project database[13]. Pink dots indicate stable phases, and the $Mg_{149}Ca$ and $Mg_{149}Al$ stable phases are not represented in this phase diagram for clarity's sake. (b) Formation enthalpy (eV/atom) as a function of Al fractions in the $(Mg_{1-x}Al_x)_2Ca$ compound. As indicated in the panel, the ground states of the $Mg_2Ca$ and $Al_2Ca$ phases are C14 and C15 structures, respectively. The global convex hull connecting two ground states is marked with a blue line. On each composition, the structure with the minimum convex hull distance is marked in the panel as a possible metastable structure.

## 7. Calculations of interfacial energies

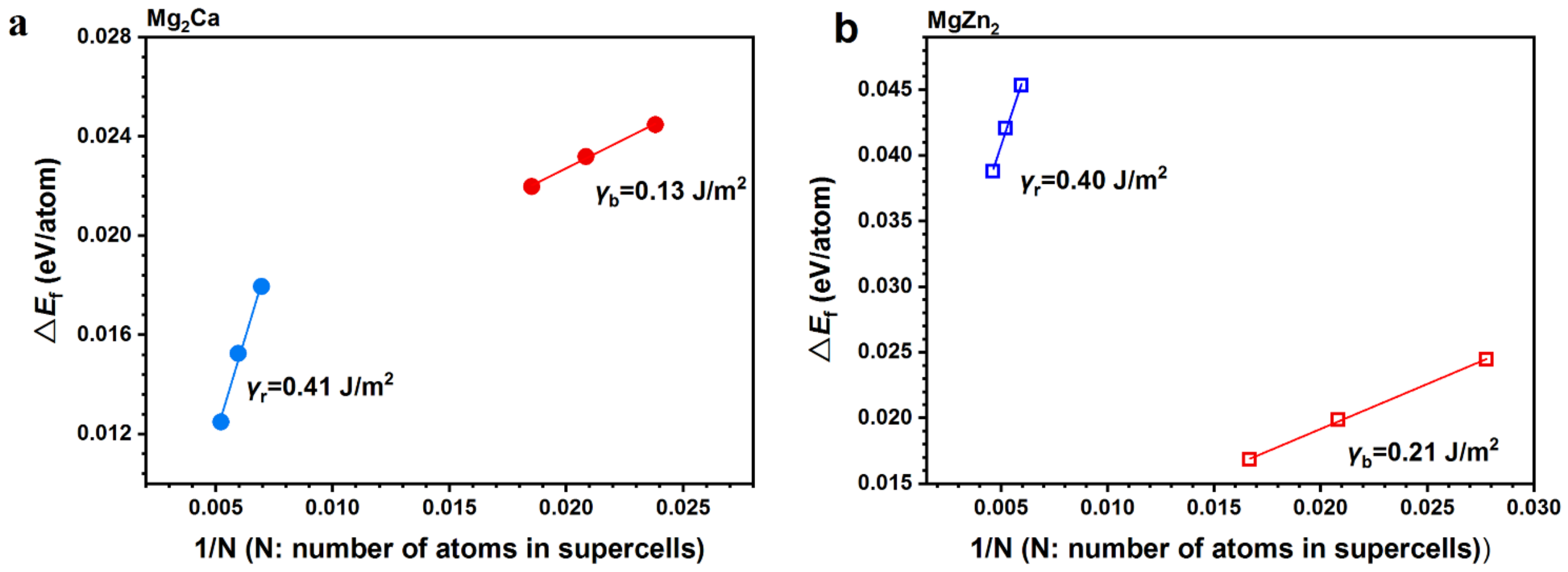

**Fig. S11.** First-principles formation energies of N-atom supercells as a function of 1/N for the $Mg_2Ca$ (a) and $MgZn_2$ (b) interface. The interfacial energies $\gamma_{b(r)}$ were extracted from the slope by Eqs. (3-4).

## 8. Supplementary References